\documentclass[superscriptaddress,showpacs,amsmath,amssymb,pra,twocolumn]{revtex4-1}

\usepackage{tikz}
\usetikzlibrary{positioning}
\tikzset{>=stealth}

\usepackage{amsfonts}
\usepackage[english]{babel}
\usepackage{braket}

\usepackage[bookmarks=false,breaklinks,bookmarksopen,bookmarksnumbered,colorlinks,linkcolor=LINKCOL,linktocpage,citecolor=CITECOL,urlcolor=URLCOL,pdfpagemode=UseOutline,pdftex]{hyperref}

\usepackage{booktabs}
\usepackage{comment}

\definecolor{TITLECOL}{rgb}{0.1,0.2,0.7} 
\definecolor{SECOL}{rgb}{0.1,0.2,0.7} 
\definecolor{CONTENTSCOL}{rgb}{0.1,0.2,0.7} 
\definecolor{SSECOL}{rgb}{0.25,0,0.48} 
\definecolor{SSSECOL}{rgb}{0.2,0.08,0.53} 
\definecolor{FINCOL}{rgb}{0.01,0.3,0.07} 

\def\coloredtitle#1{\title{\textcolor{TITLECOL}{#1}}} 
\def\coloredauthor#1{\author{\textcolor{CITECOL}{#1}}} 

\definecolor{URLCOL}{rgb}{0,0.17,0.43} 
\definecolor{LINKCOL}{rgb}{0.05,0.4,0} 
\definecolor{CITECOL}{rgb}{0.35,0,0.48} 

\usepackage{natbib}
\def\bea{\begin{eqnarray}}
\def\eea{\end{eqnarray}}
\def\ben{\begin{equation}}
\def\een{\end{equation}}
\def\benu{\begin{enumerate}}
\def\enu{\end{enumerate}}

\def\bei{\begin{itemize}}
\def\eei{\end{itemize}}
\def\beit{\begin{itemize}}
\def\eit{\end{itemize}}
\def\benu{\begin{enumerate}}
\def\enu{\end{enumerate}}

\def\hiel{\mathcal{H}_{\text{el}}}

\begin{document}

\coloredtitle{Continuous vibronic symmetries in Jahn-Teller models}

\coloredauthor{Raphael F. Ribeiro}
\affiliation{Department of Chemistry and Biochemistry, University of California San Diego, La Jolla, CA 92093}
\coloredauthor{Joel Yuen-Zhou}
\affiliation{Department of Chemistry and Biochemistry, University of California San Diego, La Jolla, CA 92093}
\date{\today}

\begin{abstract}
We develop a systematic study of Jahn-Teller (JT) models with continuous symmetries by exploring their algebraic properties. The compact symmetric spaces corresponding to JT models carrying a Lie group symmetry are identified, and their invariants used to reduce their adiabatic potential energy surfaces into orbit spaces. Each orbit consists of a set of JT distorted molecular structures with equal adiabatic electronic spectrum. Molecular motion may be decomposed into pseudorotational and radial. The former preserves the orbit, while the latter maps an orbit into another. The dimensionality and topology of the internal space of each orbit depends on the number of degenerate states in its adiabatic electronic spectra. Furthermore, qualitatively different pseudorotational modes occur in orbits of different types. We also provide a simple proof that the electronic spectrum for the space of JT minimum-energy structures (trough) displays a universality predicted by the epikernel principle. This result is in turn used to prove the topological equivalence between bosonic (fermionic) JT troughs and real (quaternionic) projective spaces, a conclusion which has outstanding physical consequences, as explained in our work. The relevance of our study for the more common case of JT systems with only discrete point group symmetry, and for generic asymmetric molecular systems with conical intersections involving more than two states is likewise discussed. In particular, we show that JT models with continuous symmetries present the simplest models of conical intersections among an arbitrary number of electronic state crossings. Our study provides a novel way to study and efficiently parametrize potential energy surfaces near conical intersections in a way which classifies the corresponding nuclear motions in a physically intuitive way.
\end{abstract}

\maketitle



 \section{Introduction}
 
\par Jahn-Teller (JT) models \cite{jahn_stability_1937, jahn_stability_1938} have been employed for the study of a variety of topics in condensed matter physics and chemistry \cite{bersuker_jahn-teller_2006, koppel_jahn-teller_2009, atanasov_vibronic_2011}. Modern works have e.g., explored JT effects in possible mechanisms for unconventional superconductivity \cite{zadik_optimized_2015}, colossal magnetoresistance \cite{millis_dynamic_1996-1},
 and single-molecule transport \cite{schultz_berry-phase_2008}. The unifying features of these phenomena are that they involve significant coupling of electronic and vibrational degrees of freedom.

\par
\begin{figure}
\includegraphics[width=0.5\textwidth]{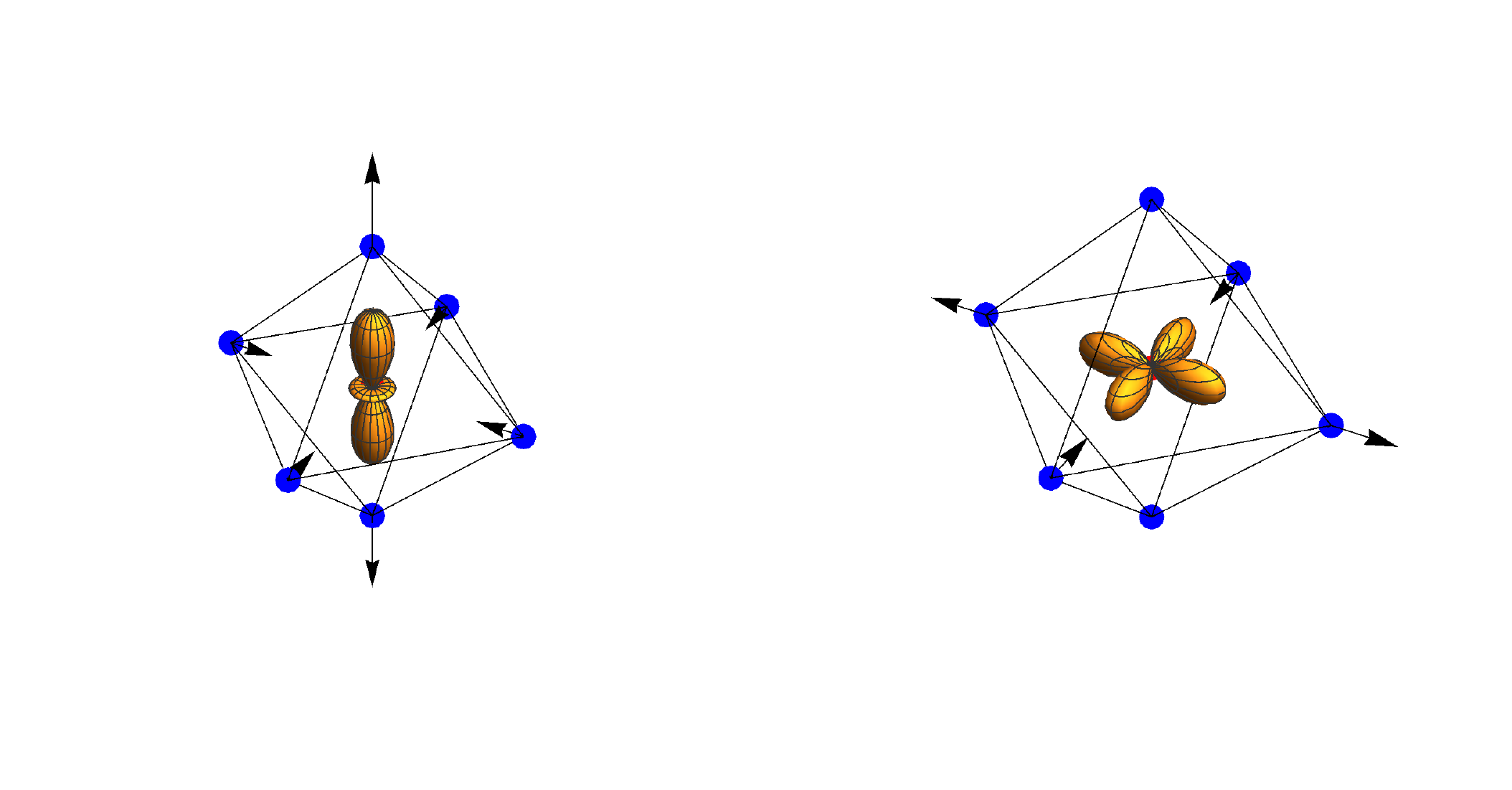}
\caption{$E_g$ vibrational modes and electronic orbitals for a system with $O_h$ symmetry}
\label{exemod}
\end{figure}

 One of the simplest JT models is that consisting of a degenerate pair of electronic states coupled to a similarly degenerate vibrational mode \cite{obrien_jahnteller_1993}, the so-called $E \otimes e$ system \cite{longuet-higgins_studies_1958} (we employ the standard convention that the irreducible representation (irrep) corresponding to the electronic multiplet is labeled by an upper-case letter, while that of the vibrations is given by a lowercase). This model has been employed, for example, to describe the distortion of Cu$^{\rm II}$ in an octahedral environment \cite{moffitt_vibronic_1957} (Fig. \ref{exemod}), and properties of triatomic systems near the equilateral configuration \cite{von_busch_unambiguous_1998}. A well-known characteristic of this system is that when only linear vibronic couplings are included, it displays circular symmetry \cite{longuet-higgins_studies_1958} (Fig. 2). The reason is the linear $E \otimes e$ Hamiltonian is invariant under simultaneous rotations of the electronic states and vibrational coordinates (see Eq. 7). The space of rotations of the plane, SO(2), is diffeomorphic to the one-dimensional circle $S^1$. Therefore, the symmetry group of linear $E \otimes e$ is \textit{continuous} (as opposed to any of the point groups, which are discrete) . This has extreme implications, e.g., there exists a continuous family of minima (trough) in the ground-state adiabatic potential energy surface (APES) (Fig. 2), the vibronic ground-state is doubly degenerate, and the vibronic (pseudo)angular momentum is quantized in odd half-integral units, thus indicating a vibronic motion with spinorial character \cite{longuet-higgins_studies_1958,bersuker_jahn-teller_2006}. All of these are surprising, as none are generic properties of finite molecular systems irrespective of the existence of point group symmetry. They are also intrinsically quantum-mechanical (even though a semiclassical treatment may lead to an effective classical Hamiltonian which describes the properties of the system when $\hbar \rightarrow 0$ \cite{zwanziger_semiclassical_1986}). 

\begin{figure}[t]
\includegraphics[width=0.35\textwidth]{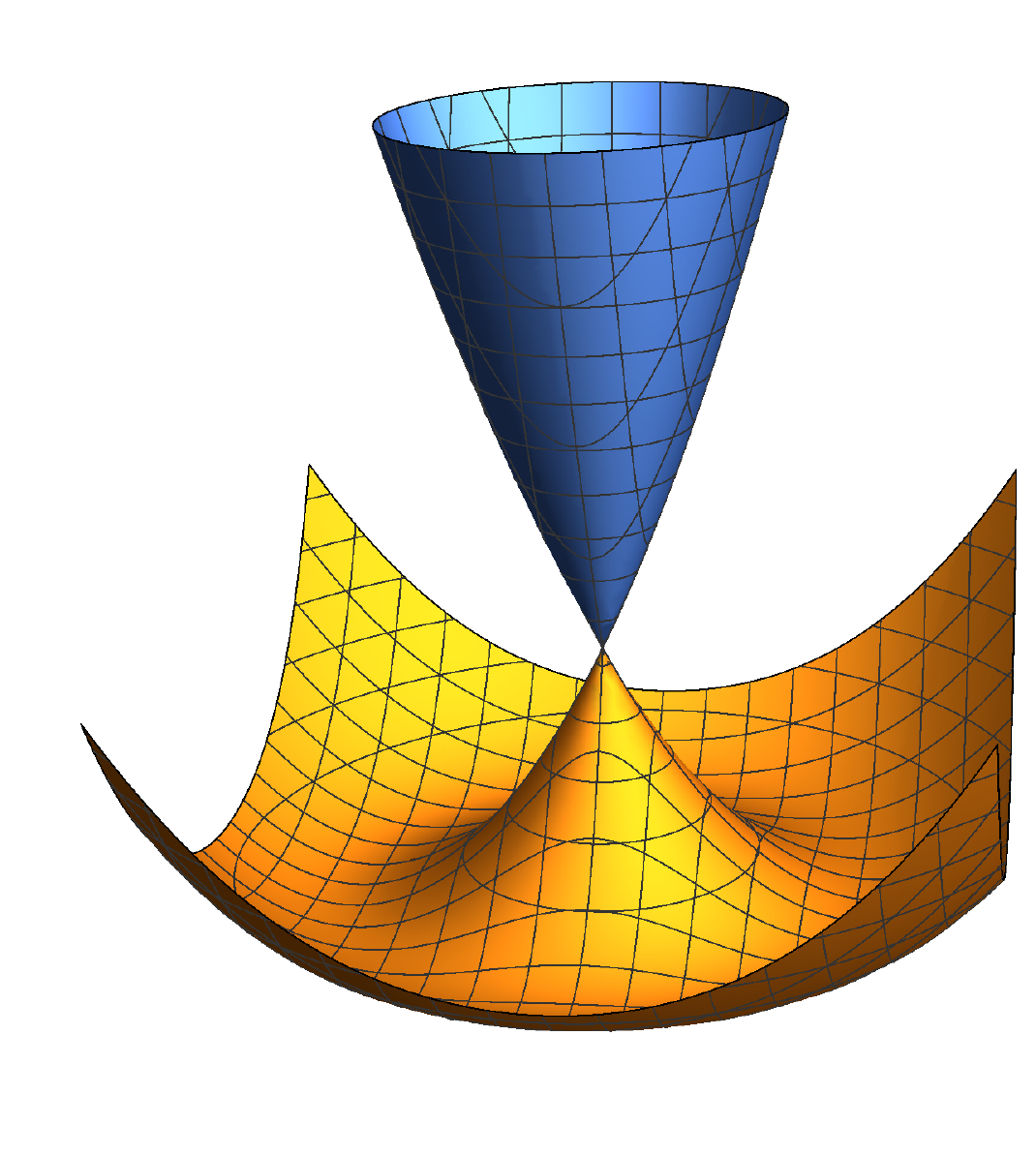}
\caption{Ground and excited-state branches of the $E \otimes e$ APES}
\label{exesurf}
\end{figure}

\par Continuous symmetries in JT systems have sometimes been described as accidental or emergent, since the molecular Hamiltonian is only constrained to be a molecular point (double) group scalar. Deeper mathematical analysis has revealed the continuous invariance properties of JT models follows from the interplay between the representations of molecular point (double) groups adopted by the electronic and nuclear degrees of freedom \cite{pooler_continuous_1978, pooler_continuous_1980,judd_group_1982}. In particular, Pooler laid out the representation theory underlying linear JT problems with continuous symmetries \cite{pooler_continuous_1978,pooler_continuous_1980}. Roughly speaking, for the latter to occur, the JT distortions need to be isotropic, and the space of electronic Hamiltonians must be equivalent to the vibrational configuration space, as irreducible representations (irreps) of a Lie group $G$. Then, the molecular Hilbert space $\mathcal{H} \equiv \mathcal{H}_{\text{vib}}\otimes \mathcal{H}_{\text{el}}$ carries a \textit{G}-action. 

\par Due to the constraints on the fundamental parameters of continuously-symmetric JT models, these have been only rarely employed to extract quantitative information about physical systems (though some examples are given below). Quadratic or non-degenerate linear vibronic couplings are known to break continuous symmetries \cite{ceulemans_a._structure_1987, bersuker_jahn-teller_2006}. For instance, in the presence of quadratic couplings, the continuous set of $E \otimes e$ minima imposed by SO(2) symmetry becomes a  discrete set separated by saddle points \cite{zwanziger_topological_1987, ceulemans_a._structure_1987, bersuker_jahn-teller_2006}. The APES associated to this model is said to be \textit{warped} (Fig. 3). However, if the distorted JT surfaces can be obtained by a gap-preserving continuous deformation (homotopy) of the electronic ground-state APES, then basic features of the continuously-symmetric JT system will remain relevant (Fig. 3) \cite{zwanziger_topological_1987}. These include the vibronic ground-state degeneracy and symmetry, as well as any other non-trivial effects originating from the existence of a Berry phase \cite{zwanziger_topological_1987, ham_berrys_1987, xie_nonadiabatic_2017}. On the other hand, if the warping generates new electronic-state intersections [esis, which include the ubiquitous conical intersections (cis) \cite{mead_determination_1979, worth_beyond_2004,truhlar_relative_2003}] on the electronic ground-state APES in regions accessible at low-energies, then the warped JT system becomes topologically different \cite{child_early_2002, zwanziger_topological_1987}. By that, we mean the symmetry and degeneracy (i.e., the irreducible representation of the molecular point group under which it transforms) of the vibronic ground-state changes. It is well-known that warping induced by introduction of quadratic vibronic couplings necessarily generates additional esis on the JT APES \cite{bersuker_jahn-teller_2006}. However, these are brought from the vibrational configuration space infinity (where they coalesce when quadratic vibronic couplings vanish) (see for example the r.h.s of Fig. 3). Thus, if the warping is weak enough, and the new esis are as a result far away, then at low-energies the molecular system will be confined to a region that excludes esis other than that which defines the JT model; in this case, homotopic invariants of the continuously-symmetric electronic ground-state APES will be preserved \cite{zwanziger_topological_1987, ham_berrys_1987}.

\begin{figure}[b]
\includegraphics[width=0.5\textwidth]{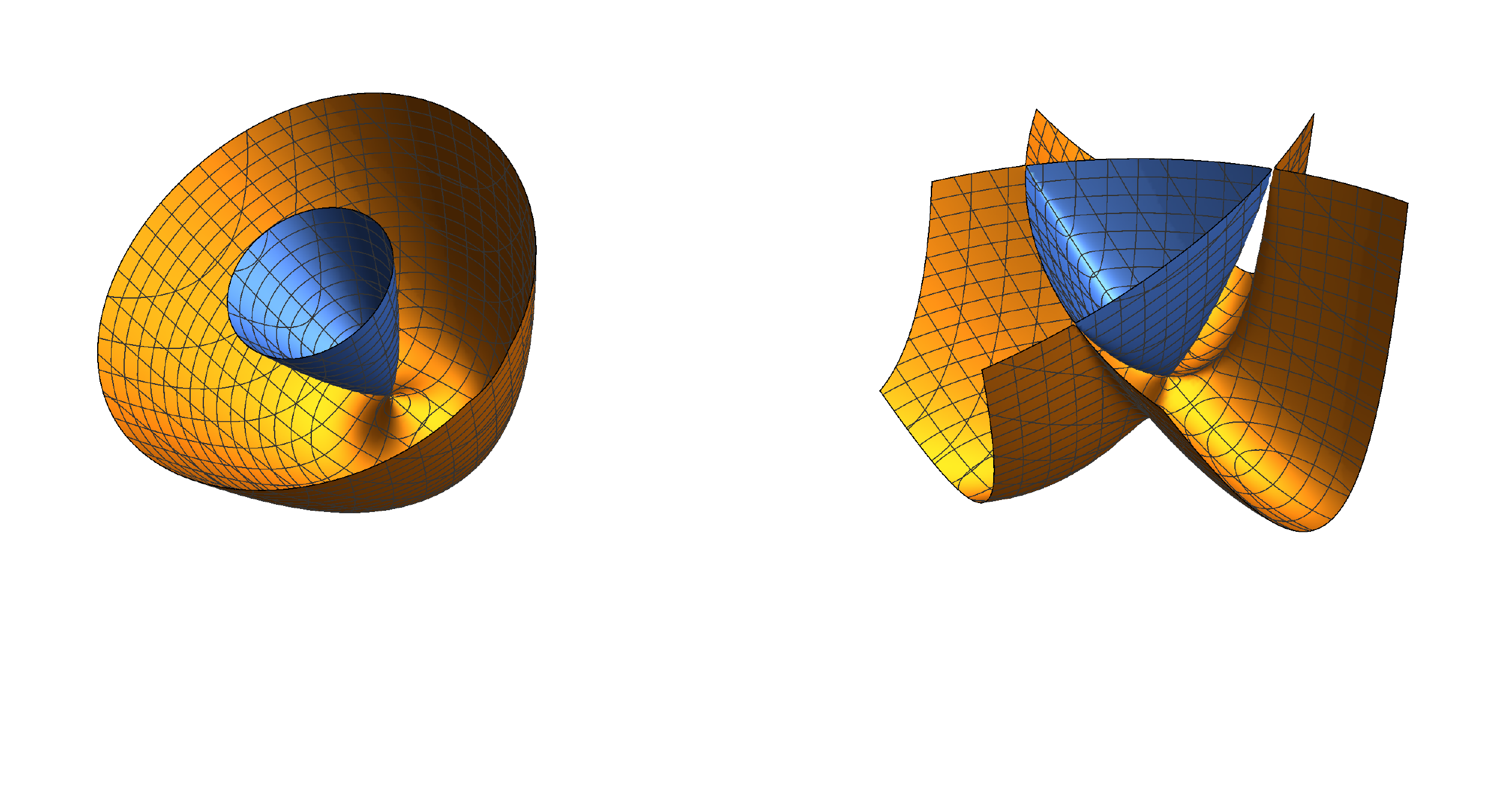}
\caption{l.h.s: weakly warped APES for $E \otimes e$; r.h.s: strongly-warped APES for the same model}
\label{exewarp}
\end{figure}

\par A less known example of the robustness of the properties of JT models with continuous symmetries was given by Markiewicz \cite{markiewicz_chaos_2001}. To understand, recall that continuous-symmetry breaking in the dynamical JT problem without quadratic or higher-order vibronic couplings will happen under one of the following three conditions: (a) different vibrational frequencies for the JT active modes irreps, but equal JT stabilization energies, (b) equal vibrational frequencies, but different JT stabilization energies, and (c) different vibrational frequencies and JT stabilization energies. In Ref. \cite{markiewicz_chaos_2001}, the linear $E \otimes (b_1 \oplus b_2)$ model (which can be understood as arising from symmetry breaking of the vibrational modes of $E \otimes e$) was employed to study case (a). This has a continuous set of ground-state minima, but pseudoangular momentum is not conserved. This happens, for the $b_1$ and $b_2$ vibrations have different frequencies. However, the average wavepacket pseudoangular momentum was verified to be quantized for different values of vibrational frequency anisotropy. While the dynamical continuous symmetry was broken, it left clear signatures. 

\par There have been also some experimental studies of JT models which benefited from an analysis based on an ideal model admitting continuous symmetries. For example, O'Brien \cite{obrien_fractional_1976} employed the SO(3)-invariant version of $T \otimes e \oplus t_2$ to investigate the spectra of F$^+$ centers (crystal vacancies occupied by holes) in CaO, and obtained quantitatively accurate absorption bandshapes. The same model also provided accurate fits to the absorption spectrum of F-centers (electron-occupied vacancies) in CsF. Another example is Pooler's and O'Brien's \cite{pooler_jahn-teller_1977} study of the $\text{P}_{1/2}^2 \rightarrow \text{P}_{3/2}^2 $ lineshapes of Tl atoms in halides. Tl is a heavy-atom and its energy levels are strongly influenced by spin-orbit coupling. The JT problem for this model is $\Gamma_8 \otimes (e\oplus t_2)$, where $\Gamma_8$ (also denoted by $G_{3/2}$) refers to the quartet double group representation of cubic systems \cite{hamermesh1962group}. With the assumption of degenerate vibronic couplings and harmonic frequencies, qualitative agreement with experimental data was also obtained.

In summary, while strong constraints need to be satisfied for a JT system to feature continuous symmetries, i.e., to carry a Lie group action, their study adds valuable insight to the generic problem of cis, as they are minimal models featuring universal properties of systems that contains them.

The connection between JT models and Lie groups has been explored before \cite{judd_lie_1974, pooler_continuous_1978, pooler_continuous_1980, judd_jahnteller_1984, apsel_berry_1992, ceulemans_a._structure_1987, ceulemans_so4_1989, ceulemans_a._and_fowler_r._jahnteller_1990, wiseman_continuous_1998, ceulemans_9_2007}. In this work we present some novel perspectives on this relationship based on the theory of \textit{symmetric spaces} \cite{helgason2001differential, magnea_introduction_2002, gilmore2012lie}. The general relevance of symmetric spaces to quantum mechanics can be illustrated with the Threefold way classification of non-relativistic Hamiltonian ensembles \cite{dyson_threefold_1962}. Dyson proved that, in the absence of anticommuting symmetries, any quantum-mechanical system belongs to either one of three symmetry classes: real orthogonal, real symplectic and complex unitary. The real orthogonal class contains all time-reversal invariant (TRI) bosonic systems, while the TRI fermionic Hamiltonians correspond to the real symplectic, and the complex unitary ensemble includes models with broken time-reversal symmetry.  Dyson's classification (and its generalization \cite{zirnbauer_symmetries_2015,ryu_topological_2010}) is the basis for the application of random matrix models in condensed matter, nuclear \cite{zirnbauer_symmetry_2004}, and chemical physics \cite{haller_statistical_1983, mukamel_fluctuations_1984}. 

Let us illustrate some of the previous remarks with the linear $E \otimes e$ system. The molecular Hamiltonian can be written as a sum of a purely vibrational part and a vibronic component, 
 \begin{equation} H(\mathbf{Q},\mathbf{P}) = H_{\text{vib}}(\mathbf{Q},\mathbf{P}) + H_{\text{JT}}(\mathbf{Q}), \mathbf{Q} \in \mathbb{R}^2, \end{equation}
 where $\mathbf{Q} = \left(Q_1,Q_2\right)$, and $\mathbf{P}=\left(P_1,P_2\right)$ are the displacements from the JT center (the molecular shape with $\mathbf{Q}=0$, hosting the electronically degenerate multiplet) and their canonically conjugate momenta, respectively (Fig. \ref{exemod}). The vibrational contribution is given by the 2D isotropic harmonic oscillator Hamiltonian,
 
 \begin{equation}  H_{\text{vib}}(\mathbf{Q},\mathbf{P}) = \frac{P_1^2}{2} + \frac{P_2^2}{2} + \frac{1}{2}\omega^2(Q_1^2+Q_2^2), \end{equation}
where $\omega$ is the vibrational frequency for the $e$ modes. The JT vibronic Hamiltonian contains the interaction between nuclear and electronic degrees of freedom. Given that $E\otimes e$ is a TRI spinless model, the time-reversal symmetry operator $T$ satisfies $T^2=1$, and we can take the electronic Hilbert space $\hiel$ to be a real vector space with basis functions invariant under $T$. Thus, the vibronic $E \otimes e$ Hamiltonian can be written as:

 \begin{equation} H_{\text{JT}}(\mathbf{Q}) = F Q  e^{-i\frac{\sigma_2}{2}\phi} \sigma_3 e^{i \frac{\sigma_2}{2}\phi}, \label{hjtexe} \end{equation}
 where $F \in \mathbb{R}-\{0\}$ is the reduced vibronic coupling constant (from the Wigner-Eckart theorem), $Q = \sqrt{Q_1^2+Q_2^2} \in \mathbb{R}^+$, $\text{tan}\phi = Q_2/Q_1$, and $\phi \in S^1$, $\sigma_j, j=1,2,3,$ are the Pauli matrices acting on $\hiel$. A change of electronic frame (basis) that preserves the reality of the electronic eigenvectors for all $\mathbf{Q}$ can be parametrized by
 
 \begin{equation} M(\theta) = e^{-i \frac{\sigma_2}{2} \theta} \in SO(2), \theta \in \mathbb{S}^1. \end{equation}
 
Let $R(\theta)$ be the rotation of vibrational configuration space defined by
 \begin{equation} \text{$R(\theta)$:}~ (Q,\phi) \mapsto (Q,\phi+\theta). \end{equation} 
 It is equivalent to the mapping  $\phi \mapsto \phi + \theta$ in Eq. \ref{hjtexe}. Hence, a change of electronic frame is equivalent to a rotation of the vibrational configuration space, 
 
  \begin{equation} M(\theta) H_{\text{JT}}\left(\mathbf{Q}\right) M^{-1}(\theta) = H_{\text{JT}}\left[R^{-1}(\theta)\mathbf{Q}\right] \label{exeinv}. \end{equation}
Hence, the JT Hamiltonian (and therefore, also the molecular) is invariant under the simultaneous action of SO(2) on the electronic and vibrational Hilbert spaces defined by
 
\begin{equation}M(\theta)  H_{\text{JT}}\left[R(\theta)\mathbf{Q}\right]M^{-1}(\theta) = H_{\text{JT}}(\mathbf{Q}). \label{cs} \end{equation}

The equation above explains the circular symmetry of Fig. 2: The SO(2) action on the space of molecular (nuclear) geometries maps a nuclear configuration into another with the same $Q$, but that is equivalent to a change of (electronic) basis of $\hiel$. Therefore, all invariant  properties of the electronic Hamiltonian are preserved (e.g., its eigenvalues) under a vibrational space rotation (pseudorotation). This explains the choice of displacement coordinates for JT distortions $(\mathbf{Q} \neq 0)$: $\phi$ is changed under SO(2) action on the vibrational configuration space (pseudorotation), while variation of $Q$ corresponds to (radial) motion that modifies the adiabatic electronic spectrum.

Equation \ref{cs} demonstrates the invariance of the linear $E \otimes e$ Hamiltonian with respect to the continuous SO(2) action on electronic and nuclear degrees of freedom. It also defines a conserved quantity by Noether's theorem \cite{jose1998classical}, the vibronic pseudoangular momentum $J_z$ (setting $\hbar = 1$ hereafter),
\begin{equation} J_z = -i\frac{\partial}{\partial \phi}+ \frac{\sigma_z}{2}, \label{psang}  \end{equation}
where the first term is a linear operator on the vibrational Hilbert space $\mathcal{H}_{\text{vib}}$, while the second acts on the electronic $\mathcal{H}_{\text{el}}$ \cite{moffitt_vibronic_1957, longuet-higgins_studies_1958}. Equation \ref{psang} is a simple, but exotic result: it implies fractionalization of the vibronic pseudoangular momentum. 

We also find that every molecular geometry of $E \otimes e$ lives in an SO(2) orbit labeled by the radial coordinate (Fig. 4). Formally, to each $Q \neq 0$ we assign the space of molecular structures
\begin{equation} O(Q) = \{ \phi \in [0,2\pi) |  R(\phi)(Q,0)^T = (Q_1,Q_2)^T\},\end{equation}
containing all $\mathbf{Q} = Q(\text{cos}\phi, \text{sin}\phi)$ with equal electronic energy spectrum. Therefore, the APES can be decomposed into a one-dimensional space of SO(2) orbits, i.e., \begin{equation} \mathbf{Q} \in \bigcup_{Q' \geq 0} O(Q'). \end{equation}
 Each point $O(Q)$ of the orbit space, with the exception of $O(0)$, has an internal space corresponding to a circle $S^1$ of molecular geometries which can be mapped into each other by free pseudorotational motion (Fig. 5). 

\begin{figure}
\includegraphics[width=0.5\textwidth]{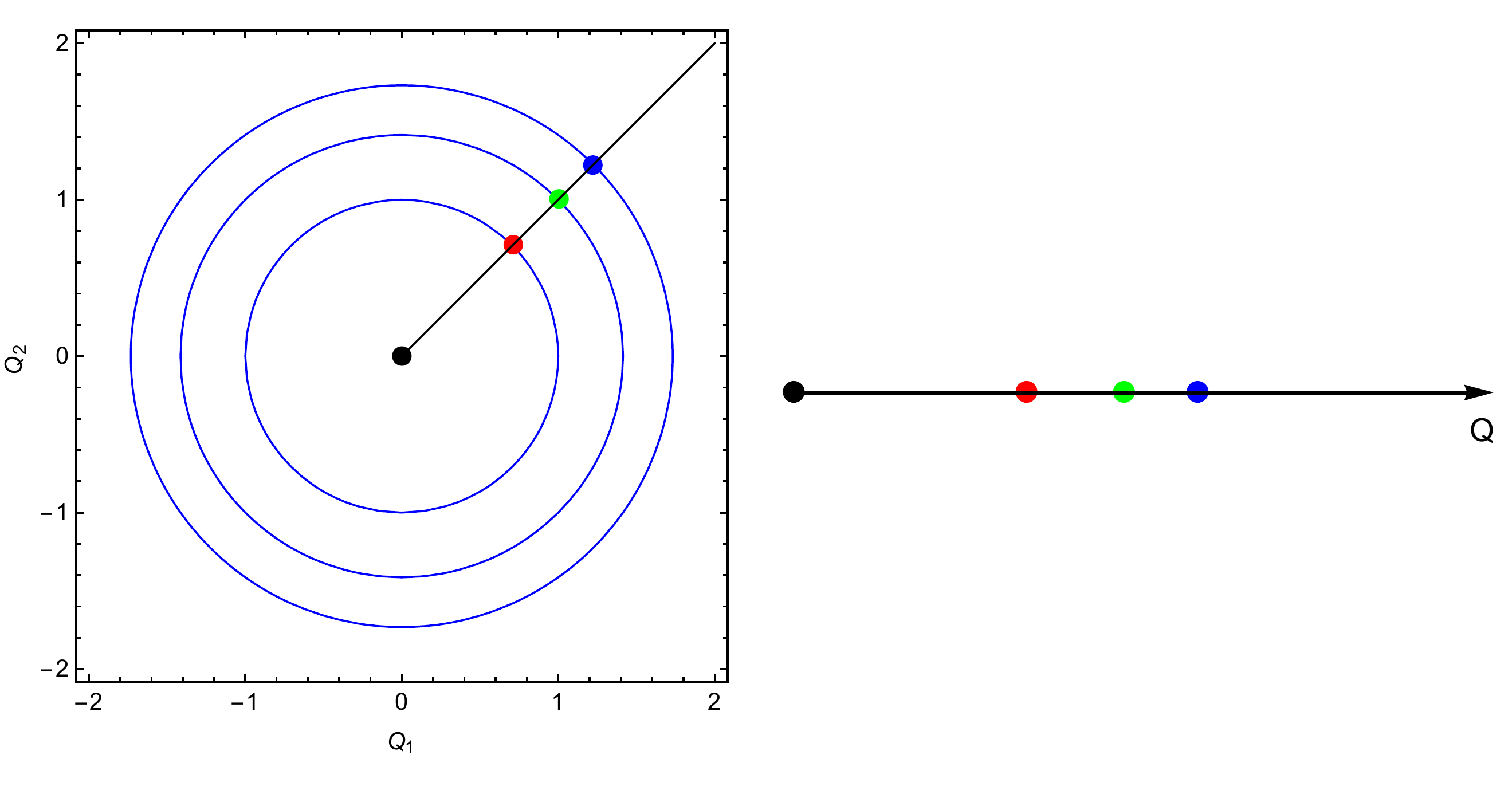}
\caption{l.h.s: $E \otimes e$ contours with equal adiabatic electronic spectrum; r.h.s: reduced orbit space}
\label{exeorb}
\end{figure}

\begin{figure}
\includegraphics[width=0.5\textwidth]{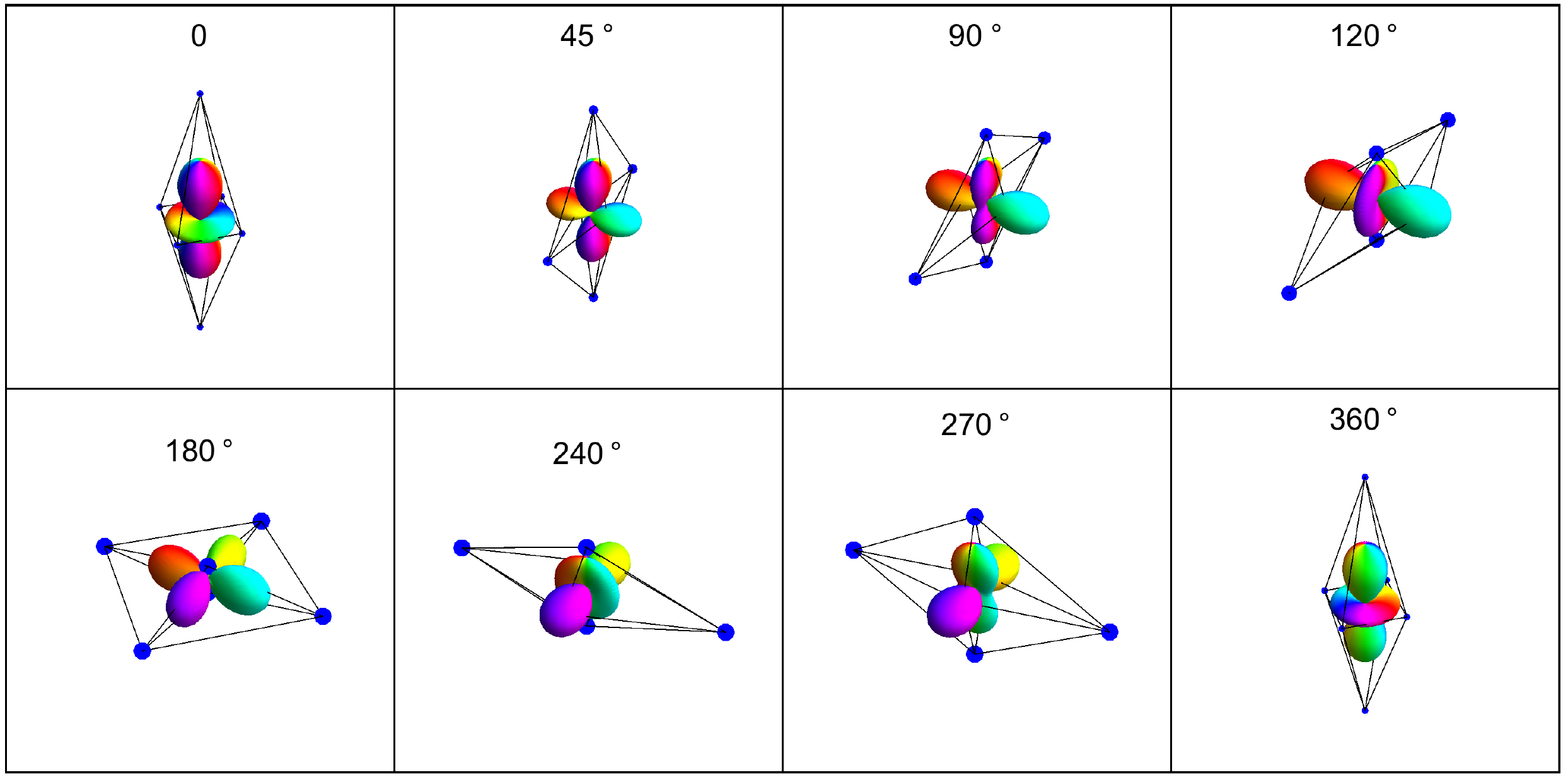}
\caption{Pseudorotational motion in vibrational and electronic configuration space (the orbital at the center is the adiabatic electronic ground-state)}
\label{exepseudo}
\end{figure}
 
These simple considerations on the SO(2) invariance of $E \otimes e$ have allowed us to quickly understand a variety of its non-trivial features. In particular, (i) the APES was simplified by decomposing it into a space of orbits of the 2D proper rotation group SO(2), (ii) the existence and meaning of radial and pseudoangular coordinates were explained, and (iii) the anomalous vibronic angular momentum conservation law of $E \otimes e$ was quickly obtained.  In more complex JT systems carrying a Lie group action, the above points are going to be generalized. For instance, there will be more than one angular and one radial coordinate, so that orbits will require more than a single number to be specified uniquely. There will also be orbits of different types with qualitatively distinct internal (pseudorotational) motion. The invariance properties of symmetric spaces will be useful in the development of these generalizations.
   
 \par This paper is organized as follows: in section II we discuss the general relationship between linear JT problems with continuous symmetries and symmetric spaces. Section III applies the framework presented in the previous to characterize local and global properties of the JT APES. In section IV we discuss the connection between the investigated models and molecular systems with accidental cis involving more than two APESs. A summary of our main results and a discussion of future directions is given in section V.      
\section{Symmetric spaces underlying JT models with continuous symmetries}

We start this section by providing a quick review of the conditions satisfied by JT systems admitting continuous symmetries. These are used to determine the symmetric spaces corresponding to each element of the set of JT models studied in this article.  We also discuss some general properties of symmetric spaces which will be useful for establishing several universal properties of JT systems carrying a Lie group action.

 Consider a JT problem defined by the coupling between electronic states belonging to the irrep $\Gamma$ of a point group $S$ ($S$ is the symmetry group of the nuclear geometry at $\mathbf{Q}=0$, the JT center) spanned by the basis vectors $\{\gamma_i\}_{i=1,2,...,|\Gamma|}$, and JT active vibrations transforming like the vectors of the (generically) reducible representation $\lambda = \oplus_{i} \Lambda_i$, where $\Lambda_i$ is a non-totally symmetric irrep of $S$ with basis $\{\lambda_{i_1},\lambda_{i_2}...\lambda_{i_{|\lambda_i|}}\}$. The electronic multiplet is only defined to belong to a single irrep, for none of the studied models contain reducible electronic representations (Wiseman showed how to generalize the theory of Ref. \cite{pooler_continuous_1980} to the case of direct sum electronic Hilbert spaces \cite{wiseman_continuous_1998}). Generalizing Eqs. $1-3$, the molecular Hamiltonian of the aforementioned system including only linear vibronic couplings is given by \cite{bersuker_jahn-teller_2006}
 
\begin{align} H(\mathbf{Q},\mathbf{P}) = \sum_{\Lambda_i}\frac{\mathbf{P}_{\Lambda_i}^2}{2m_{\Lambda_i}} + \sum_{\Lambda_i}\frac{1}{2} m_{\Lambda_i} \omega_{\Lambda_i}^2 \mathbf{Q}_{\Lambda_i}^2 + H_{\text{JT}}(\mathbf{Q}),  \label{ham} \end{align}

\begin{align} H_{\text{JT}}(\mathbf{Q}) = \sum_{\Lambda_i}F_{\Lambda_i \Gamma} \mathbf{Q}_{\Lambda_i}\cdot\mathbf{V}_{\Lambda_i \Gamma}, \end{align}
where $\mathbf{Q}_{\Lambda_i} = (Q_{\Lambda_i \lambda_{i_1}}, Q_{\Lambda_i \lambda_{i_2}}...,Q_{\Lambda_i \lambda_{i_{|\Lambda_i|}}})$, (with similar notation for $\mathbf{P}_{\Lambda_i}$), $m_{\Lambda_i}$ and $\omega_{\Lambda_i}$ are the mass and harmonic frequency of the nuclear displacements $\Lambda_i$, $F_{\Lambda_i}$ is the reduced vibronic coupling constant associated with the same modes (from the Wigner-Eckart theorem), and $\mathbf{V}_{\Lambda_i\Gamma}$ is a vector of Clebsch-Gordan matrices, i.e.,  $\mathbf{V}_{\Lambda_i\Gamma} = (V_{\Lambda_i \lambda_{i_1}\Gamma}, V_{\Lambda_i \lambda_{i_2}\Gamma}...,V_{\Lambda_i \lambda_{i_{|\Lambda_i|}}\Gamma})^T$, with e.g.,

\begin{equation} V_{\Lambda_i \lambda_{i_1}\Gamma} =\sum_{\gamma_k \in \Gamma} \sum_{\gamma_l \in \Gamma} \braket{\Gamma \gamma_k |\Lambda_i \lambda_{i_1} \Gamma \gamma_l} \ket{\Gamma \gamma_k}\bra{\Gamma \gamma_l}. \end{equation}

Pooler has given the formal theory underlying the construction of molecular JT Hamiltonians with continuous symmetries, i.e., invariant under the action of a Lie group \cite{pooler_continuous_1978,pooler_continuous_1980}. First, it was shown that the space of electronic tensor operators living in $\Gamma \otimes \Gamma$ can be decomposed into even and odd subspaces, which we will denote by $\mathcal{M}$ and $\mathcal{P}$, respectively. The operators in $\mathcal{M}$ belong to the symmetric part of $\Gamma \otimes \Gamma-A_1$ (where $A_1$ is the totally symmetric irrep) if the system is spinless (we discuss the fermionic case later). The odd live in the antisymmetric. Therefore, $\mathcal{P}$ is  a Lie subalgebra of $\Gamma \otimes \Gamma - A_1$ (since the commutator of antisymmetric matrices is antisymmetric). Conversely, the commutator of even operators is an odd element, so the latter do not span a Lie algebra.

$\mathcal{P}$ generates the $|\Gamma|$-dimensional special orthogonal group, SO$(|\Gamma|)$. If $\mathcal{M}$ is isomorphic to an irrep $\overline{\Lambda}$ of SO$(|\Gamma|)$, then $H(\mathbf{Q},\mathbf{P})$ is invariant under the action of the same group on both electronic and nuclear degrees of freedom, if and only if the vibrational configuration space can also be embedded in (i.e., transforms like) the irrep $\overline{\Lambda}$ of SO$(|\Gamma|)$ \cite{pooler_continuous_1980}.

In other words, sufficient conditions for invariance under continuous transformations of electronic and vibrational degrees of freedom in JT models are that: (a) only linear vibronic couplings are non-vanishing, and (b) the space of traceless electronic Hamiltonians and the vibrational configuration space are isomorphic as irreps of a Lie group $G$. More formally, let End$(\Lambda)$ correspond to the space of linear transformations of the vibrational configuration space $\Lambda$ while End $[\text{Herm}(\Gamma \otimes \Gamma - A_1)]$ is the space of linear operators acting on the vector space generated by the Hermitian electronic matrices of $\Gamma \otimes \Gamma - A_1$. Lie group representations for each of these spaces are defined by maps $\text{$R$:} ~G \rightarrow \text{End}(\Lambda)$ and $\text{$U$:} ~G \rightarrow \text{End}[~\text{Herm}(\Gamma \otimes \Gamma)-A_1]$, respectively. Condition (b) is encapsulated by the following diagram,

\begin{center}
\begin{tikzpicture}
    \node (G) at (0,0) {$G$};
    \node[right=of G] (F) {$\text{End}[\text{Herm}(\Gamma\otimes\Gamma-1)]$};
    \node[below=of G] (M) {$\text{End}(\Lambda)$};
    \draw[->] (G)--(F) node [midway,above] {$U$};
    \draw[->] (G)--(M) node [midway,left] {$R$};
    \draw[<->] (F) -- (M);
\end{tikzpicture}
\end{center}
where the two sided arrow means that there is a one-to-one relationship between the representations defined by $U$ and $R$.

 The  above conditions require that the electronic multiplet $\Gamma$ is equally coupled to every JT active mode, while $m_{\Lambda_i} = m$ and $\omega_{\Lambda_i} = \omega_\Lambda$ for all $\Lambda_i \subset \Lambda$. Physically, they imply equal JT stabilization energies for molecular distortions along each $\Lambda_i$, and (pseudo)rotational symmetry in the purely vibrational part of $H(\mathbf{Q},\mathbf{P})$, respectively.

The embeddings of the vibrational and electronic point group irreps $\Lambda = \oplus \lambda_i$ and $\Gamma$ into Lie group irreps will not be subject of future discussion. The interested reader may consult Refs. \cite{pooler_continuous_1978, pooler_continuous_1980,chancey1997jahn} for details. Henceforth we take the conditions for continuous invariance of JT Hamiltonians to be fulfilled and study its basic consequences. For the sake of simplicity, we will not make a distinction between the continuous and point group irreps anymore, so that from now on $\Lambda$ and $\Gamma$ correspond to the appropriate continuous group irreps of SO$(N)$ or USp$(2N)$ depending on the total spin of the considered electronic JT multiplet.

The relations satisfied by the $\hiel$ operators in $\cal{P}$ and $\cal{M}$ can be summarized by

\begin{equation} [\cal{M},\cal{M}] = \cal{P}, ~[\cal{P},\cal{M}] = \cal{M},~ [\cal{P}, \cal{P}]= \cal{P}. \label{csym} \end{equation}
They imply that $\cal{L} = \cal{M} \oplus \cal{P}$ is a Lie algebra \cite{gilmore2012lie}. The latter corresponds to the tangent space of a Lie group $L$ at the identity. The odd operators define a subalgebra $\cal{P}$. They generate the Lie subgroup $P \subset L$. Importantly, the electronic matrices in $\mathcal{M}$ define tangent vectors for the \textit{symmetric space} $L/P$ \cite{helgason2001differential,magnea_introduction_2002, gilmore2012lie}. In fact, the algebraic definition of symmetric spaces is encapsulated by the relations shown in Eq. \ref{csym}. Some of their relevant basic properties will be discussed below. 

The requirement that the time-reversal operator $T$ satisfies $T^2=1$ is important, since it implies the existence of a basis where the electronic Hamiltonian matrix elements are \textit{real}. Thus, only the electronic tensor operators of $\mathcal{L} \subset sl(|\Gamma|,\mathbb{R})$ (the space of traceless real matrices)  which are in the symmetric part of $\Gamma \otimes \Gamma - A_1$ lead to allowed quantum Hamiltonians. Hence, $H_{\text{JT}}(\mathbf{Q})$ belongs to $\mathcal{M} = \mathcal{L}/\mathcal{P}$. As noted before, the \textit{antisymmetric} electronic tensor operators generate the special orthogonal groups SO$(|\Gamma|)$. Its generators define the Lie algebra $so(N)$: the space of infinitesimal rotations of $\hiel$. Thus, $so(N)$ acts on the operator $H_{\text{JT}}(\mathbf{Q})$ by infinitesimal rotations. As we show below, while the defined $so(N)$ action occurs on $\hiel$, it is equivalent to infinitesimal rotation of the internal displacements $\mathbf{Q} \rightarrow \mathbf{Q} + \delta \mathbf{Q}$ in JT models with continuous symmetries.

In the spinless case, $\mathcal{M} \oplus \mathcal{P}$ spans either (a) $sl(N,\mathbb{R})$ (see below for the $T^2=-1$ case), or (b) one of its proper subsets \cite{pooler_continuous_1978}.  Examples of class (a) are the Lie group invariant formulations of $E \otimes e$ \cite{longuet-higgins_studies_1958}, $T \otimes (e \oplus t_2)$ \cite{obrien_dynamic_1969-2}, $T \otimes h$ \cite{khlopin_jahn-teller_1978}, $G \otimes (g\oplus h)$ \cite{ceulemans_so4_1989},  $H \otimes (g\oplus 2h)$ \cite{ceulemans_a._and_fowler_r._jahnteller_1990, de_los_rios_dynamical_1996, chancey1997jahn}. These include JT active distortions representing all different ways to split the electronic degeneracy at $\mathbf{Q}=0$. In other words, every allowed non-totally symmetric electronic coupling mediated by vibrational modes is included in the vibronic Hamiltonian. In these models, the continuous symmetry is \textit{maximal}, since the symmetry of the molecular Hamiltonian cannot be increased without increasing the size of $\hiel$. Conversely, class (b) contains $H \otimes h$ \cite{khlopin_jahn-teller_1978,ceulemans_a._and_fowler_r._jahnteller_1990}, $H\otimes(g\oplus h)$ \cite{ceulemans_a._and_fowler_r._jahnteller_1990}, etc \cite{pooler_continuous_1978}. Their vibronic coupling matrices only span a proper subset of $sl(N,\mathbb{R})/so(N,\mathbb{R})$, and the molecular Hamiltonian symmetry can be increased by including the remaining JT active vibrational modes. Thus, the latter models do not include one or more of the possible JT active distortions. In this paper we focus only on the JT models with maximal continuous symmetries (see Table 1), for they share a variety of deeply related properties. Thus, from now on, anytime we mention JT models carrying a Lie group action, it should be understood that we are referring to models in class (a), unless otherwise noted.
 
The electronic Hamiltonian can only be constructed with the even tensor operators $V_\alpha \in \mathcal{M}$. Therefore, the constraints given above imply the most general JT Hamiltonian for a $\Gamma \otimes \Lambda$ system with a continuous symmetry can be written as:

\begin{equation}H = \frac{\mathbf{P}_{\Lambda}^2}{2m} +\frac{1}{2}k_\Lambda \mathbf{Q}_\Lambda^2 +F_\Lambda
\sum_{\alpha \in \cal{M}} Q_\alpha V_\alpha, \label{hamcs} \end{equation}
where $k_{\Lambda} = m \omega_\Lambda^2$. Note the matrices $V_\alpha \subset \mathcal{M}$ and the $so(|\Gamma|)$ generators are both real and traceless, so their direct sum is the algebra $sl(|\Gamma|,\mathbb{R})$, i.e., $\mathcal{L}=sl(|\Gamma|,\mathbb{R})$. As a consequence, the underlying symmetric spaces for the spinless class of JT Hamiltonians with continuous symmetries is SL$(|\Gamma|,\mathbb{R})$/SO$(|\Gamma|)$ (or SU$(|\Gamma|)$/SO$(|\Gamma|)$ if we take the skew-hermitian $\hiel$ time-evolution generators $iV_{\alpha}$ as the generators of the symmetric space).

A fundamental property of symmetric spaces is their rank \cite{gilmore2012lie,helgason2001differential}. It provides the number of \textit{anisotropic} spatial directions on these manifolds. Physically, the rank of a given JT model is equal to the number of linearly independent nuclear distortions unrelated by SO($|\Gamma|$) (or USp$(|\Gamma|)$ in the fermionic case) transformations. For instance, in $E \otimes e$ we can define a radial and an angular coordinate for the vibrational configuration space (when $\mathbf{Q} \neq 0$), but only geometries with the same radial coordinate $Q$ are related by nuclear pseudorotation. Therefore, in the space of time-evolution operators of $E \otimes e$, there exists a single direction which is anisotropic. This conforms with the fact that the rank of SU(2)/SO(2) is one \cite{gilmore2012lie,helgason2001differential}.  

Another view on the rank of a symmetric space is that it specifies the dimensionality of its maximal abelian subalgebra (also commonly denoted Cartan subalgebra) $\mathcal{C}$ \cite{helgason2001differential}.  By a theorem of symmetric spaces theory, any element $\mathfrak{m}$ of $\mathcal{L}/\mathcal{P}$, is related to a member $\mathfrak{m}_C$ of the Cartan subalgebra $\mathcal{C}$ by conjugation with an element of the Lie group $P$ (with tangent space at the identity $\mathcal{P}$)\cite{helgason2001differential}, 
\begin{equation} \mathfrak{m} = g \mathfrak{m}_C g^{-1}, g \in P, \mathfrak{m}_C \in \mathcal{C}. \label{carconj} \end{equation}
The rank of the symmetric space SL$(|\Gamma|,\mathbb{R})/$SO$(|\Gamma|)$ (or SU$(|\Gamma|)/$SO$(|\Gamma|$)), is equal to $|\Gamma|-1$ \cite{gilmore2012lie}.  

We can always take the Cartan subalgebra of $sl(|\Gamma|,\mathbb{R}|)/so(|\Gamma|)$ to be given by $|\Gamma|-1$ linearly independent traceless diagonal matrices of $sl(|\Gamma|,\mathbb{R})$ \cite{gilmore2012lie}. In this case, Eq. \ref{carconj} is the simple statement that any real traceless symmetric matrix $\mathfrak{m}$ can be diagonalized by an orthogonal transformation. We note that conjugation preserves commutation relations, so the choice of Cartan subalgebra is not unique. A set of diagonal matrices may be the simplest, but it is not the only.

It follows that the electronic part of the molecular Hamiltonian  (Eq. \ref{hamcs}) can be rewritten as

\begin{equation} \begin{split} H_{\text{JT}}(\mathbf{Q}) = & F_{\Lambda}\sum_{\alpha \in \mathcal{M}} Q_\alpha V_\alpha  \\ = &~ U(\mathbf{Q}) \left(F_\Lambda\sum_{\alpha_C \in \mathcal{M}_C} Q_{\alpha_C} V_{\alpha_C}\right)U^{-1}(\mathbf{Q}), \\~ U(\mathbf{Q}) \in \text{SO}(|\Gamma|),\label{hjts} \end{split}\end{equation}
where $\mathcal{M}_C \subset \mathcal{M}$ is a choice of Cartan subalgebra for $\mathcal{M} \equiv sl(N,\mathbb{R})/so(N)$.
The assumed duality between traceless Hermitian electronic tensor operators and JT distorted geometries automatically implies that there exists an embedding of SO$(|\Gamma|)$ in SO$(|\Lambda|$), since proper transformations of the vibrational configuration space are described by the latter group. Hence, it follows that there exists a continuous map, $\text{$R$:}~ \text{SO}(|\Gamma|) \rightarrow \text{SO}(|\Lambda|)$, such that

\begin{align} \sum_{\alpha_C \in \mathcal{M}_C}U(\mathbf{Q}) Q_{\alpha_C} V_{\alpha_C}U^{-1}(\mathbf{Q}) =  \sum_\alpha [R^{-1}(\beta)\mathbf{Q}_C]_{\alpha} V_\alpha,  \label{adjr}  \end{align}
where $\beta$ denotes an SO($|\Lambda|$) coordinate, $R(\beta)$ specifies the SO$(|\Lambda|)$ rotation satisfying $R^{-1}(\beta) \mathbf{Q}_C = \mathbf{Q}$, and 
$\mathbf{Q}_C$ is the nuclear displacement with nonvanishing components only along the directions which are dual to the Cartan subalgebra matrices $V_{\alpha_C}$. Eq. \ref{adjr} states that in JT models with continuous symmetries, a rotation of the electronic frame is equivalent to a rotation of the space of JT distorted structures. It is the generalization of Eq. \ref{exeinv}.

In cases where spin-orbit coupling is strong and time-reversal symmetry is implemented by $T$ satisfying $T^2=-1$, the JT active modes live in the \textit{antisymmetric} part of the $\Gamma \otimes \Gamma - A_1$ space, where $\Gamma$ is now a spinorial double group irrep \cite{altmann2005rotations, jahn_stability_1938}. The electronic Hilbert space for a fixed geometry can thus be given the structure of a quaternionic vector space \cite{avron_chern_1989}, the unitary transformations of which are implemented by the unitary symplectic group U$(|\Gamma|/2,\mathbb{H})\cong$ USp$(|\Gamma|)$ \cite{gilmore2012lie}. The molecular Hamiltonian is invariant under the simultaneous action of the symplectic group on the electronic and nuclear degrees of freedom whenever the symmetric part of $\Gamma \otimes \Gamma$ generates the $usp(|\Gamma|)$ algebra, the traceless antisymmetric part lives in $su^*(|\Gamma|)/usp(|\Gamma|)$ [$su^*(|\Gamma|) \cong sl(|\Gamma|/2,\mathbb{H})$ is the space of traceless $|\Gamma|\times|\Gamma|$ complex matrices invariant under conjugation by an antiunitary operator $T$ satisfying $T^2=-1$, or simply the space of traceless $|\Gamma|/2\times|\Gamma|/2$ matrices with quaternionic entries \cite{gilmore2012lie}], and the vibrational degrees of freedom transform in the same irrep of USp$(|\Gamma|)$ as the traceless antisymmetric electronic tensor operators \cite{pooler_continuous_1978, pooler_continuous_1980}. Therefore, in cases with strong spin-orbit coupling, the underlying symmetric spaces for JT models carrying a Lie group action consist of SU*$(|\Gamma|)$/USp$(|\Gamma|)$ [or SU$(|\Gamma|)$/USp$(|\Gamma|)$, if instead of considering Hamiltonians as generators, we consider the corresponding time-evolution generators $iH$].  A list of all JT models carrying a Lie group action encompassed by our study is given in Table I, along with their corresponding symmetric spaces, and examples of point groups giving rise to these models.

\begin{table}
\caption{JT models with continuous symmetries investigated in this paper. The last column gives examples of discrete symmetry groups originating each model \cite{ceulemans2013group}. S* refers to the double group obtained from a point group S.}
\centering
\begin{tabular}{|c|c|c|c|c|}
\hline
JT model & $T^2$ & Symmetric Space & Examples  \\
\hline
$E \otimes e$ \cite{longuet-higgins_studies_1958} & 1 & SU(2)/SO(2) & $C_{3v}, O_h$ \\
\hline 
$T \otimes (e\oplus t_2)$ \cite{obrien_dynamic_1969-2} & 1 & SU(3)/SO(3) & $Td, O_h, I, I_h$\\
\hline
$G \otimes (g\oplus h)$ \cite{ceulemans_so4_1989} & 1 & SU(4)/SO(4) & $I, I_h$\\
\hline
$H \otimes (g\oplus 2h)$ \cite{ceulemans_a._and_fowler_r._jahnteller_1990} & 1 & SU(5)/SO(5) & $I, I_h$ \\
\hline
$\Gamma_8 (G_{3/2}) \otimes (e\oplus t_2)$ \cite{pooler_jahn-teller_1977} & $-1$ & SU(4)/USp(4) & $Td^*, O_h^*, I_h^*$  \\
\hline
$\Gamma_9 (I_{5/2}) \otimes (g\oplus 2h)$ \cite{ceulemans_9_2007} & $-1$ & SU(6)/USp(6) &$ I^*,I_h^*$ \\
\hline
\end{tabular}
\end{table}

\section{Local and global structure of JT orbits}

We now take advantage of the framework introduced in the previous section to quickly obtain insight into the APESs of JT models carrying a Lie group action. As we show in Sec. IV., the results which we present here also have some significance for the treatment of dynamics in the neighborhood of electronic degeneracies of generic molecular systems beyond JT models. Therefore, they display a degree of universality which is perhaps unanticipated in view of the constraints imposed by the continuous symmetries of the investigated models.

\subsection{Symmetry-adapted coordinates and orbit spaces}
Consider first the problem of choosing coordinates for vibrational motion, which are adapted to the invariance properties of the underlying symmetric space. The advantages of a suitable choice of coordinate system cannot be overstated. For instance, in $E \otimes e$, the identification of radial and angular coordinates provides automatic insight into the dynamical JT problem. They are also useful in describing the warped ground-state APES \cite{ceulemans_a._structure_1987}.

Coordinates adapted to the circular symmetry of the SO(2) $E \otimes e$ model can be trivially obtained. However, this is not the case for higher-dimensional models with SO($N), N > 2$ or USp$(2N)$ symmetries. The situation is ameliorated if only the space of minima of the JT APES is of interest, for the ground-state APES is homemorphic with the real (if $T^2=1$ \cite{ceulemans_a._structure_1987}) or quaternionic (when $T^2=-1$) projective space (see Secs III.B.2 and III.B.3). The real projective space ($\mathbb{R}P^N$) can be obtained from the sphere $S^N$ by identifying antipodal points of the latter \cite{nakahara2003geometry}. Therefore, hyperspherical coordinates are a natural choice for the study of vibrational motion on the spinless ground-state trough \cite{ceulemans_so4_1989, ceulemans_a._and_fowler_r._jahnteller_1990} of the continuously-symmetric JT models. While profitable in discerning properties of the space of minima of the electronic ground-state APES, this approach gives no insight into the motion which is transverse to the extremal APES, nor does it explain the spectral flow of the Born-Oppenheimer Hamiltonian in non-stationary regions of the molecular internal space. It does not take full advantage of the invariance properties of the systems studied here.

\begin{figure}
\includegraphics[width=0.5\textwidth]{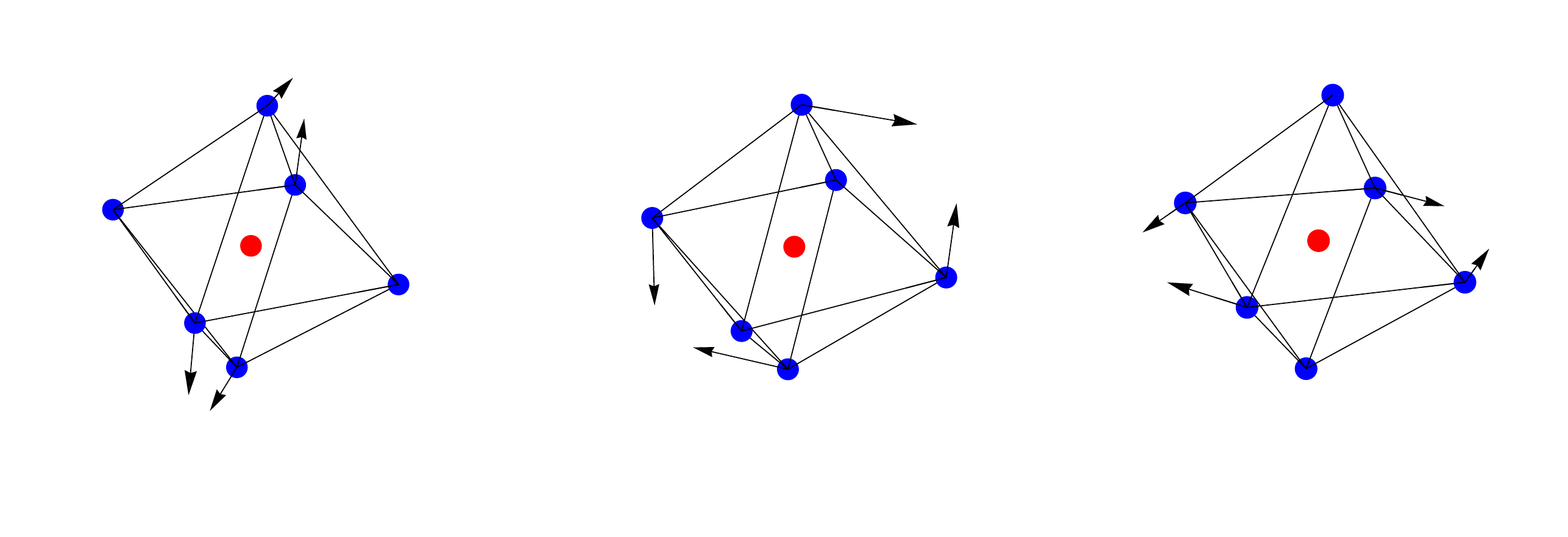} 
\caption{$T_{2g}$ vibrational modes for a system with $O_h$ symmetry}
\label{t2mod}
\end{figure}

We will employ the isomorphism between traceless electronic Hamiltonian operators and vibrational displacements as irreps of a Lie group $G$, to choose vibrational coordinates adapted to the corresponding SO($N$) (or USp($2N$)) action. Consider first the specific case of the cubic JT problem $T_1 \otimes (e\oplus t_2)$ \cite{obrien_dynamic_1969-2}, or equivalently the icosahedral $T \otimes h$ (the generalization to more complex models will be made later) \cite{chancey1997jahn}.  The continuous symmetry of this system in the presence of degenerate couplings was originally investigated by O'Brien \cite{obrien_dynamic_1969-2, obrien_jahn-teller_1971}. The JT active modes are characterized by the vector $\mathbf{Q}= (Q_\theta, Q_\epsilon, Q_{xy},Q_{zx},Q_{yz})$ (see Figs \ref{exemod} and \ref{t2mod}). A global diabatic basis for $\hiel$ is defined by $\mathcal{H}_{\text{el}} = \text{span}\{\ket{x},\ket{y},\ket{z}\}$ (which may be though of as $p_x,p_y$ and $p_z$ atomic orbitals). The traceless Hermitian electronic tensor operators are in the symmetric part of $T_1 \otimes T_1 - A_1$ and will be labeled by $V_{\alpha}$, $\alpha \in \{\theta, \epsilon, xy, zx, yz\}$,

\begin{equation} \begin{split} &V_{\theta} = \left( \begin{array}{ccc}
\frac{1}{2} & 0 & 0 \\
0 & \frac{1}{2}& 0 \\
0& 0 &-1 \end{array} \right), V_{\epsilon}=\left( \begin{array}{ccc}
-\frac{\sqrt{3}}{2} & 0 & 0 \\
0 & \frac{\sqrt{3}}{2} & 0 \\
0& 0 &0 \end{array} \right), \\ &
(V_{ij})_{ab} = -\frac{\sqrt{3}}{2}\left(\delta_{ia}\delta_{jb} + \delta_{ja}\delta_{ib}\right), \label{matet2} \end{split}\end{equation}
where $i,j \in \{x,y,z\}, ~i \neq j$, $\delta_{ab} = 1$ if $a=b$ and 0 otherwise.

The active JT distortions and the space of electronic Hamiltonians are isomorphic to the space of quadrupoles \cite{avron_chern_1989} [the $\textbf{2}$ irrep of SO(3)], which is also isomorphic to the tangent space of SL$(3,\mathbb{R})$/SO$(3)$ at the identity, i.e., $sl(3,\mathbb{R})/so(3)$.  Its Cartan subalgebra is two-dimensional. It may be taken to be  for example, $\text{span}\{V_{\epsilon},V_{\theta}\}$ or alternatively $\text{span}\{V_{\theta},V_{xy}\}$. Thus, there exists two SO(3) invariants in the universal enveloping algebra of $sl(3,\mathbb{R})/so(3,\mathbb{R})$ \cite{helgason2001differential}. From the SO(3)-equivalence between Hermitian electronic tensor operators (minus the identity)  and vibrational displacements, it follows that there exists two functions $r_1(\mathbf{Q})$ and $r_2(\mathbf{Q})$ of the nuclear degrees of freedom which are invariant under the action of SO(3). They will be denoted \textit{radial coordinates} in analogy with the rotational invariance of the radius of an $N$-dimensional ball.  Let $r_1$ denote the quadratic SO(3) invariant of $sl(3,\mathbb{R})/so(3)$, $r_1(\mathbf{Q}) = |\mathbf{Q}|^2 = Q$, and $r_2(\mathbf{Q})$ be the cubic. The latter may be obtained from the secular determinant of $H_{\text{JT}}(\mathbf{Q})$ \cite{helgason2001differential}, or as the totally symmetric SO(3) irrep $\mathbf{0}$ in the decomposition of the tensor product $\mathbf{2}\otimes \mathbf{2} \otimes \mathbf{2}$, since the cubic invariant is a homogeneous polynomial of third degree in the vibrational coordinates\cite{judd_jahnteller_1984}. We provide a simple derivation of $r_2(\mathbf{Q})$ below.
 
This simple analysis indicates that each molecular geometry of $T \otimes (e\oplus t_2)$ belongs to an SO(3) orbit characterized by two real   numbers, $r_1$ and $r_2$. We denote a specific orbit of this model by $O(r_1,r_2)$. It only includes molecular geometries with the same electronic spectrum, since $r_1$ and $r_2$ are SO(3) invariants. It follows that given a representative $\mathbf{Q} \in O(r_1,r_2)$, any element $\mathbf{Q}' \in O(r_1,r_2)$ may be obtained by the SO(3) action on $\mathbf{Q}$, this rotation in turn corresponding to the electronic Hilbert space SO(3) transformation $U(\mathbf{Q}',\mathbf{Q})$ satisfying $U(\mathbf{Q},\mathbf{Q}') H_{\text{JT}}(\mathbf{Q}) U^{-1}(\mathbf{Q},\mathbf{Q}') = H_{\text{JT}}(\mathbf{Q}')$. Hence, we verify that the 5-dimensional APES of $T \otimes (e\oplus t_2)$ may be reduced into a 2-dimensional orbit space.

The remaining three vibrational degrees of freedom are expected to parametrize the internal space of the orbits $O(r_1,r_2)$. Thus, SO(3)-adapted coordinates for the molecular displacements are given by $\mathbf{Q}= \mathbf{Q}(r_1,r_2,\beta,\alpha,\phi), ~\forall ~r_1 \in \mathbb{R}-\{0\}, r_2 \in \mathbb{R}$, where $(\beta,\alpha, \phi)$ parametrize SO(3) rotations. They can be understood as Euler angles in the space of JT active distorted structures \cite{judd_jahnteller_1984, auerbach_electron-vibron_1994}.  We employ the $zyz$ convention for the parametrization of SO(3) elements,

\begin{equation} U(\beta,\alpha,\phi) = e^{-i \phi J_z} e^{-i\alpha J_y}e^{-i \beta J_z}, \label{so3rot}\end{equation}
where $U(\beta,\alpha,\phi)$ acts on the \textbf{1} (vector) irrep of SO(3), the $J_i$ are its Hermitian generators, $(J_i)_{jk} = i \epsilon_{ijk}$, $\phi, \beta \in [0,2\pi)$, and $\alpha \in [0,\pi]$.

In order to understand the internal space of each orbit, note that the JT distortion-induced electronic multiplet splitting can only happen in two different ways: either $H_{\text{JT}}(\mathbf{Q})$ (with $\mathbf{Q} \neq 0$) has three distinct eigenvalues or two degenerate and a single non-degenerate. The matrices representing the vibronic coupling due to $t_2$ or $e$ displacements demonstrate this point. $H_{\text{JT}}(\mathbf{Q})$ is related by an orthogonal transformation to an element of the Cartan subalgebra of $sl(3,\mathbb{R})/so(3)$. In particular, if we choose $V_{\theta}$ and $V_\epsilon$ to be a basis for the latter, we find from Eq. \ref{hjts},

\begin{equation} H_{\text{JT}}(\mathbf{Q}) = F_{v}U(\beta,\alpha,\phi)\left(Q_c^{\epsilon} V_{\epsilon}+Q_c^\theta V_\theta \right) U^{-1} (\beta,\alpha,\phi), \end{equation}
\begin{widetext}
\begin{equation} H_{\text{JT}}(\mathbf{Q}) = 
-F_v |\mathbf{Q}| 
U(\beta,\alpha,\phi)
\left( \begin{array}{ccc}
\text{cos}[\gamma(\mathbf{Q}_c)-\frac{2\pi}{3}]& 0& 0 \\
0 & \text{cos}[\gamma(\mathbf{Q}_c)+\frac{2\pi}{3}] & 0 \\
0& 0 &\text{cos}[\gamma(\mathbf{Q}_c)] \end{array} \right)
U^{-1}(\beta,\alpha,\phi), \label{quadham} \end{equation} \label{hsim} \end{widetext}
where $F_v$ denotes the vibronic coupling constant, $\mathbf{Q}_c = (Q_c^{\theta},Q_c^\epsilon,0,0,0)$ defines an element of the Cartan subalgebra of $sl(3,\mathbb{R})/so(3)$ specifying the eigenvalues of the JT Hamiltonian for a given geometry $\mathbf{Q} = R^{-1} \mathbf{Q}_c, R \in$ SO(3), and $\gamma(\mathbf{Q}) = \text{tan}^{-1}(Q_c^\epsilon/Q_c^\theta)$. Note $U(\beta,\alpha,\phi)$ is a function only of Euler angles since $r_1(\mathbf{Q})$ and $r_2(\mathbf{Q})$ are SO(3)-invariants. These considerations drastically simplify the task of finding the coordinate mapping $\mathbf{Q} \mapsto r_2(\mathbf{Q})$ (the cubic invariant of $sl(3,\mathbb{R})/so(3)$). It is proportional to the $O(t^3)$ term of $\text{det}[H_{\text{JT}}(\mathbf{Q})-t \lambda]$ \cite{helgason2001differential}, which may now be easily calculated from Eq. \ref{quadham},

\begin{equation} r_2(\mathbf{Q}) = Q^3 \text{cos}[3\gamma(\mathbf{Q}_c)].\end{equation}

 For any $\mathbf{Q}$, $H_{\text{JT}}(\mathbf{Q})$ is related by an orthogonal transformation to a \textit{unique} $H_{\text{JT}}(\mathbf{Q}_c)$ with $\gamma \in [0,\pi/3]$\cite{avron_chern_1989}, so $r_2 \in (-\infty,\infty)$. 

 According to the above, motion along the $\mathbf{Q_\theta}$ and $\mathbf{Q}_\epsilon$ directions provides the only distinct possibilities for the splitting of the electronic multiplet when $Q\neq 0$. This may also be seen by checking that the eigenvalues of $V_{ij}$ are equal to those of $V_{\epsilon}$, which in turn are different from those of $V_\theta$. The orbit obtained by SO(3) action on $V_\theta$  defines a 2D subspace of the vibrational configuration space, since $[J_z, V_\theta] = 0$. In more detail, suppose $\mathbf{Q}= Q \mathbf{e}_\theta = (Q,0,0,0,0)$. Then, $\mathbf{Q}_c =\mathbf{Q}$, and
 
\begin{equation} U(0,0,\phi) Q V_\theta U^{-1}(0,0,\phi) = Q V_\theta. \end{equation} 
This equation implies the space of molecular geometries obtained by pseudorotation of the structure defined by $\mathbf{Q} = Q \mathbf{e_\theta}$ is specified by only two of the three SO(3) parameters. In fact, the orbits containing $Q \mathbf{e}_\theta=(1,0,0,0,0)$, or $-Q\mathbf{e}_\theta$, can be parametrized uniquely by a 3D rotation axis $\mathbf{n} = \mathbf{n}(\phi,\alpha) = \mathbf{n}(\pi+\phi,\pi-\alpha)$ (see Eq. \ref{so3rot}). Conversely, rotations of the non-degenerate Cartan subalgebra basis vector $V_{\epsilon}$ give a 3D subspace. If we take $r_1(\mathbf{Q}) = |\mathbf{Q}|^2 = 1$, the corresponding SO(3) orbit will be either two or three-dimensional depending on the value of $r_2(\mathbf{Q})$. For instance, while $\mathbf{Q} =  \mathbf{e}_\theta$ lives in a 2D orbit, $\mathbf{e}_{\epsilon}=(0,1,0,0,0)$ belongs to a 3D. Note $r_2(\pm\mathbf{e}_\theta) = \pm 1$, while $r_2(\mathbf{e}_{\epsilon}) = 0$. In fact, the $T \otimes (e\oplus t_2)$ JT Hamiltonian contains an adiabatic electronically degenerate doublet iff $\gamma = 0$ or $\gamma = \pi/3$, i.e., for $r_2 = \pm Q$.  For any other values of $\gamma \in [0,\pi/3]$, the corresponding electronic JT Hamiltonian belongs to a 3D orbit of SO(3).
 
Our analysis also reveals the topology of each type of orbit. The 2D orbits are parametrized by a 3D rotation axis. Thus, each 2D orbit is isomorphic to $\mathbb{R}P^2$. The 3D orbits are copies of SO(3)/V $\cong \mathbb{R}P^3/V$ \cite{nakahara2003geometry}, where $V = \{U(0,0,0), U(0,0,\pi), U(0,\pi,0)\, U(\pi/2,\pi,\pi/2)\}$. The latter follows from the fact that every diagonal matrix is left invariant under conjugation by an element of the subgroup $V$. Therefore, electronic Hilbert space transformations by elements of $V$ correspond to trivial action on the vibrational configuration space.
  
 The qualitative analysis carried to this point indicates the APES of the SO(3)-invariant $T\otimes (e\oplus t_2)$ can be decomposed into a 2D orbit space. Only two types of orbits exist in agreement with the two linearly independent ways of lifting the degeneracy of triply-degenerate electronic states. 
 
By applying Eqs. 19-21 to \ref{quadham}, we  obtain the SO(3)-adapted parametrization of JT distortions \cite{obrien_dynamic_1969-2},
$\mathbf{Q} =  \mathbf{Q}(Q,\gamma, \beta,\alpha,\phi)$,
  
 \begin{widetext}
\begin{eqnarray} &Q_{\epsilon} =Q \left[\frac{\sqrt{3}}{2}\text{cos}(\gamma)\text{cos}(2\phi) \text{sin}^2(\alpha) + \frac{1}{2}\text{sin}(\gamma)[[1+\text{cos}^2(\alpha)]\text{cos}(2\phi)\text{cos}(2\beta)-\text{cos}(\alpha)\text{sin}(2\beta)\text{sin}(2\phi)] \right]\nonumber \\
&Q_{\theta} = Q\left[\frac{1}{2}\text{cos}(\gamma)(\text{cos}^2(\theta)-1) + \frac{\sqrt{3}}{2}\text{sin}(\gamma)\text{sin}^2(\alpha)\text{cos}(2\beta)\right]\nonumber \\
 &Q_{xy}  = Q \left[\frac{\sqrt{3}}{2}\text{cos}(\gamma) \text{sin}^2 (\alpha)\text{sin}(2\phi) + \frac{1}{2}\text{sin}(\gamma)[(1+\text{cos}^2\alpha)\text{sin}(2\phi)\text{cos}(2\beta)+2\text{cos}(\alpha) \text{cos}(2\phi)\text{sin}(2\beta)] \right] \nonumber \\ 
& Q_{zx} = Q \left[ \frac{\sqrt{3}}{2}\text{cos}(\gamma)\text{sin}(2\alpha)\text{cos}(\phi)+ \frac{1}{2}\text{sin}(\gamma)[-\text{sin}(2\alpha) \text{cos}(2\beta) \text{cos}(\phi)+\text{sin}(\alpha)\text{sin}(2\beta) \text{sin}(\phi)]  \right] \nonumber \\
 &Q_{yz} = Q\left[\frac{\sqrt{3}}{2}\text{cos}(\gamma)\text{sin}(2\alpha)\text{sin}(\phi) + \frac{1}{2}\text{sin}(\gamma)[-\text{sin}(\alpha)\text{sin}(2\phi)\text{sin}(2\beta) -\text{sin}(2\alpha)\text{sin}(\phi)\text{cos}(2\beta)]\right]. 
 \end{eqnarray} \label{coords} \end{widetext}
 
This parametrization was first given (in the context of JT models) by O'Brien \cite{obrien_jahn-teller_1971}. It was of paramount importance for studies of the fullerene $T \otimes h$  JT model \cite{auerbach_vibrations_1994, chancey1997jahn}. The derivation above gives a simple systematic method for its construction, which can be applied to the other, more complex JT systems with continuous symmetries (see below).
 
The coordinate transformations described by Eq. 25 allow a quick visualization of some of the features of the two types of orbits we discussed: $\gamma = 0~\text{or}~\pi/3 \implies \mathbf{Q} = \mathbf{Q}(Q,\alpha,\phi)$, thus characterizing a 2D orbit parametrized by $\alpha \in [0,\pi]$ and $\phi \in [0,2\pi)$. Pseudorotational motion in the 3D orbits (where $\gamma \neq 0$ or $\pi/3$) is specified by the Euler angles $\beta,\alpha,\phi$.
  
The above considerations can be generalized to the other JT models carrying a Lie group action (see Table 1), including the fermionic cases with strong spin-orbit coupling (in the following, simply make the replacements $sl(n,\mathbb{R}) \rightarrow sl(n,\mathbb{H}) \cong su^*(2n,\mathbb{C})$, and $so(n) \rightarrow usp(2n)$ and all results will remain valid for the fermionic models). The maximal commuting subalgebra of $sl(n,\mathbb{R})/so(n)$ is $n-1$-dimensional \cite{gilmore2012lie}. Therefore, there exists a set of $n-1$ anisotropic directions in the corresponding JT APES. Each characterizes an independent radial molecular distortion $r_i(\mathbf{Q})$, which specifies a particular type of splitting of the electronic spectrum. The values taken by the radial displacements $r_1,r_2,...,r_{n-1}$ specify SO$(N)$ orbits labeled by $O(r_1,...,r_{n-1})$. Thus, the orbit space of an SO($N)$-invariant model contains $n-1$ dimensions. Each point of the orbit space is an orbit belonging to one of a finite number of types determined by the number of degenerate electronic multiplets. As shown explicitly for the  $T \otimes (e \oplus t_2)$ case, different orbits may have internal spaces with varying number of degrees of freedom. While the adiabatic electronic spectrum is completely specified by the radial displacements, a complete description of the molecular geometry requires specification of angular coordinates corresponding to the pseudorotational motion on each orbit.  These can be parametrized by SO$(n)$ variables (e.g., we used Euler angles in the SO(3) case). For instance, in $G \otimes (g\oplus h)$, the symmetric space is SU(4)/SO(4). It has 3 radial and 6 angular coordinates which can be constructed from the set of JT active vibrations $g\oplus h$. SU(5)/SO(5) arises in the $H \otimes (g \oplus 2h)$ model, which includes 4 radial and 10 angular coordinates. The radial degrees of freedom can be directly obtained in all cases by looking at the coefficients of the invariant polynomials produced by the JT Hamiltonian secular determinant, $\text{det}[H_{\text{JT}}(\mathbf{Q}) - t\lambda]$. Angular displacements adapted to a simple description of pseudorotation can be retrieved by a straightforward generalization of the procedure carried for $T_1 \otimes (e\oplus t_2)$. For example, in the case of $G \otimes (g\oplus h)$, a molecular geometry with non-degenerate electronic spectrum will be characterized by \textit{at most} two sets of Euler angles, as $so(4) \cong so(3) \oplus so(3)$ \cite{gilmore2012lie}. 

\subsection{APES troughs and higher-energy orbits}

In the previous subsection, we explored the invariant properties of JT models carrying a continuous group action to obtain symmetry-adapted coordinates, which fully take advantage of the high-symmetry of the studied models. As we have shown, these coordinates are directly related to the decomposition of the JT active vibrational motion into pseudorotational and radial. Pseudorotation preserves the adiabatic electronic spectrum, while radial motion nontrivially modifies the latter. Therefore the APES can be reduced into an orbit space, where each orbit defines a subspace with constant electronic spectrum. Motion on the latter is generated by pseudoangular momentum. This provides significant simplification of the dynamical JT problem, since the orbit space has a smaller number of degrees of freedom than the vibrational configuration space. In this subsection, we explore further properties of JT models to classify their troughs and higher-energy orbits according to the qualitatively distinct splittings of the corresponding adiabatic electronic multiplets. In Sec. III.B.1, we show that the electronic ground-state minima of all JT models with maximal continuous symmetries have a distinct spectral signature. This result is utilized in sections III.B.2, where we prove that the ground-state trough of each spinless (fermionic) JT model with continuous symmetries is topologically equivalent to a real (quaternionic) projective space. This equivalence is of vital importance to the discussion of the Berry phase in Sec. III.C. We provide comments on the properties of higher-energy orbits in Sec. III.B.3.

\subsubsection{Ground-state trough for JT models with maximal continuous symmetries}

Here we generalize an argument of O'Brien \cite{obrien_dynamic_1969-2} to show that for JT models with maximal continuous symmetries, the equilibrium molecular geometries live in the SO($N$) orbit $\mathcal{O}$ with adiabatic electronic spectrum given by
\begin{align}\text{spec}\left[\hat{H}_{\text{JT}}(\mathbf{Q})\right] = \{x(Q),x(Q),...,-(N-1)x(Q)\}, \nonumber \\ \mathbf{Q} \in \mathcal{O}, x(Q) > 0.  \label{speceq}\end{align} 
Suppose the vibrational states live in the $\Lambda$ irrep of a point group $S$, and let the adiabatic electronic ground-state and its corresponding JT APES be denoted by $\ket{\psi_0(\mathbf{Q})}$ and $V^{(0)}(\mathbf{Q})$, respectively. The latter has contributions from the JT stabilization energy $E_{\rm{JT}}^{(0)}(\mathbf{Q})$ and the harmonic potential energy $k_\Lambda \mathbf{Q}^2/2$, i.e.,
 \begin{equation} V^{(0)}(\mathbf{Q}) = E_{\rm{JT}}^{(0)}(\mathbf{Q}) + \frac{1}{2}k_{\Lambda} \mathbf{Q}^2, ~E_{\rm{JT}}^{(0)}(\mathbf{Q}) = \text{inf} ~\hat{H}_{\rm{JT}}(\mathbf{Q}). \label{vjt} \end{equation}
Equilibrium molecular geometries $\mathbf{Q}_0$ satisfy
 
 \begin{align} & \left(\nabla_{\mathbf{Q}}V^{(0)}\right)(\mathbf{Q}_0) = 0, \nonumber\\ &\sum_{i,j}\delta Q_{\Lambda i} \delta Q_{\Lambda j} \left(\nabla_{Q_{\Lambda i}}\nabla_{Q_{\Lambda j}}V^{(0)}\right)(\mathbf{Q}_0)> 0,  \end{align} 
 where $\delta Q_{\Lambda i} = \left(\mathbf{Q} - \mathbf{Q}_0\right)_{\Lambda i}$. With the help of Eq. \ref{vjt}, the first condition can be restated as
 \begin{equation} \frac{\partial E_{\rm{JT}}^{(0)}}{\partial Q_{\Lambda \lambda}}(\mathbf{Q})\bigg|_{\mathbf{Q}=\mathbf{Q}_0} = -k_{\Lambda} (Q_0)_{\Lambda \lambda},~\forall~\lambda \in \{1,2,...,|\Lambda|\}. \end{equation}
 
By the continuous symmetry assumption there is a continuous set of solutions to the above, all of which are related by rotation of the coordinates via an SO($N$) action. Because $Q$ is preserved by this action, all minima may be generated by rotations of $\mathbf{Q}_i = Q(0...,0,1,0..,0)$, where $i$ is the nonvanishing component.
 
Let a basis for linearly independent traceless symmetric electronic tensor operators be denoted by $M_{\Lambda \lambda}$. Then, $\hat{H}_{\rm{JT}}(\mathbf{Q})$ can be expressed by

\begin{equation} \hat{H}_{\rm{JT}}(\mathbf{Q}) = F_\Lambda\sum_{\lambda=1}^{|\Lambda|} M_{\Lambda \lambda} Q_{\Lambda \lambda}. \end{equation} 
The presence of SO($N$) symmetry implies the existence of a choice of electronic states and vibrational coordinates, such that, $N-1$ of the matrices $M_{\Lambda \lambda}$ are diagonal, while the rest are symmetric with only two non-vanishing (off-diagonal) elements \cite{pfeifer2003lie}. A general element of the former and latter sets will be called $D_{\lambda}$ and $O_{\lambda}$, respectively. The diagonal matrices can be parametrized in the following manner \cite{pfeifer2003lie}: 
\begin{align} 
&D_1 = \{1,-1,0,...,0\}, D_2 = \frac{1}{\sqrt{3}}\{1,1,-2,0,...0\},  ...,\nonumber \\ &D_{N-1} = \sqrt{\frac{2}{N(N-1)}} \{1,1,...,-N+1\} \label{dmat},
\end{align}
where each set contains the entries of the corresponding diagonal matrix. We take the entries of the $O_\Lambda$ to be equal to $-1$ (for an example, see Eq. \ref{matet2}). For instance, the matrices of Eq. \ref{matet2} can be put in this form by the mapping $V_\Lambda \mapsto (2/\sqrt{3}) V_\Lambda$. We expect $D_{N-1}$ to dictate the splitting of the JT degeneracy at the electronic ground-state minima, since a molecular distortion along the direction corresponding to $D_{N-1} $ maximally stabilizes the electronic ground-state at the expense of destabilization of all higher-energy states. In fact, this insight agrees with a prediction based on the epikernel principle \cite{ceulemans_symmetry_1984, ceulemans_a._structure_1987,ceulemans_epikernel_1989}, which states that the extrema of the JT ground-state APES are likely to be encountered in regions of the vibrational configuration space which preserve a large subgroup of the symmetry group at $\mathbf{Q}=0$. The maximal subgroup of SO($N$) is SO($N-1$). The latter is indeed preserved under a molecular distortion along the displacement $\mathbf{Q}_{N-1}$ corresponding to the JT splitting defined by $D_{N-1}$.

Let $M_{\Lambda N-1} = D_{\Lambda,N-1}$, and consider the ground-state JT APES at $\mathbf{Q} = \mathbf{Q}_{N-1} \equiv Q(0,0,...,1)$,
 
 \begin{equation} V^{(0)}(\mathbf{Q}_{N-1})  = -F_\Lambda \sqrt{\frac{2(N-1)}{N}}Q + \frac{1}{2} k_{\Lambda} Q^2. \end{equation}
The condition for $\mathbf{Q}_{N-1}$ to be an extremum of $V^{(0)}$ gives

\def\Vi{V^{(i)}(\mathbf{Q}_{N-1})}

\begin{align} &Q = \frac{F_\Lambda}{k_\Lambda} \sqrt{\frac{2(N-1)}{N}}, \nonumber \\\ & V^{(0)}(\mathbf{Q}_{N-1}) = -\frac{F_\Lambda^2}{k_{\Lambda}}\frac{N-1}{N}, \nonumber \\
&\Vi =  \frac{F_{\Lambda}^2}{k_\Lambda}\frac{N+1}{N}, i \in \{1,2,...,N-1\}. \label{eqcon} \end{align}
We will use second-order perturbation theory to study the behavior of the JT ground-state APES in a neighborhood of $\mathbf{Q}_{N-1}$ with radius $|\delta\mathbf{Q}| = \epsilon \rightarrow 0^+$. The 0th-order distorted APES is defined by
\begin{equation} H_0(\mathbf{Q}_{N-1}+\delta \mathbf{Q}) = H_{\rm{JT}}(\mathbf{Q}_{N-1}) + \frac{1}{2}k_{\Lambda}\left(\mathbf{Q}_{N-1} +\delta \mathbf{Q}\right)^2, \end{equation} while the vibronic perturbation due to infinitesimal motion $\delta\mathbf{Q}$ can be written as:

\begin{equation} H'(\delta\mathbf{Q}) = F_\Lambda \sum_{\lambda}\delta Q_{\Lambda \lambda} M_{\Lambda \lambda}. \end{equation}
Thus, to second-order in perturbation theory the electronic ground-state APES is given by:

\begin{widetext}
\begin{equation} V^{(0)}(\mathbf{Q}_{N-1} +\delta\mathbf{Q}) = V^{(0)}(\mathbf{Q}_{N-1})+k_\Lambda \mathbf{Q}_{N-1} \cdot \delta \mathbf{Q} +  \frac{1}{2}k_{\Lambda} (\delta \mathbf{Q})^2 + F_\Lambda\sum_{\lambda=1}^{|\Lambda|} [M_{\Lambda \lambda}]_{NN} \delta Q_{\Lambda \lambda} +  \sum_{\lambda=1}^{|\Lambda|} \sum_{i=1}^{N-1} \frac{F_\Lambda^2\left[M_{\Lambda \lambda}\right]_{iN}^2}{V^{(0)}(\mathbf{Q}_{N-1})-V^{(i)}(\mathbf{Q}_{N-1})}\delta Q_{\Lambda\lambda}^2.\end{equation}
\end{widetext}
The above can be simplified by using Eq. \ref{eqcon} to obtain the relation $V^{(0)}(\mathbf{Q}_{N-1})-\Vi = -2F_{\Lambda}^2/k_\Lambda$, and Eq. \ref{dmat} for $[M_{\Lambda \lambda}]_{NN}$, whence

\begin{align} & \delta V^{(0)} (\mathbf{Q}_{N-1}) = k_\Lambda Q \delta Q_{\Lambda |\Lambda|} +  \frac{1}{2}k_{\Lambda} (\delta \mathbf{Q})^2 \nonumber\\ &-F_\Lambda \sqrt{\frac{2(N-1)}{N}}\delta Q_{\Lambda |\Lambda|} - k_\Lambda\sum_{\lambda=1}^{|\Lambda|} \sum_{i=1}^{N-1} \frac{\left[M_{\Lambda \lambda}\right]_{iN}^2}{2}\delta Q_{\Lambda\lambda}^2,
\end{align}
where $\delta V^{(0)} (\mathbf{Q}_{N-1}) = V^{(0)}(\mathbf{Q}_{N-1} +\delta\mathbf{Q}) - V^{(0)}(\mathbf{Q}_{N-1})$. As expected, the first and third terms cancel so that the ground-state JT APES is given by

\begin{equation} \delta V^{(0)} (\mathbf{Q}_{N-1})  = \frac{k_\Lambda}{2}\sum_{\lambda=1}^{|\Lambda|} \left[1 - \sum_{i=1}^{N-1}\left[M_{\Lambda \lambda}\right]_{iN}^2 \right] \delta Q_{\Lambda\lambda}^2. \label{inteq}\end{equation}
To proceed we recall the general behavior of $\left[M_{\Lambda \lambda}\right]_{iN}$. Only the symmetric matrices in $O_\Lambda$ with off-diagonal elements $[M]_{iN} = -1$ contribute to the second sum of the above equation. There exists $N-1$ symmetric matrices satisfying this condition. Their contributions to $\delta V^{(0)} (\mathbf{Q}_{N-1})$ will be cancelled by the first term of Eq. \ref{inteq}. Let $M_{\Lambda a}$ denote the diagonal matrices $D_i$ and the symmetric matrices with vanishing $iN$ entry. Then, it follows that 

\begin{equation}  \delta V^{(0)} (\mathbf{Q}_{N-1}) =  \sum_{a} \frac{1}{2}k_{\Lambda} \delta Q_{\Lambda a}^2. \end{equation}
Hence, we find that the SO($N$) orbit $\mathcal{O}$ defines a continuous set of absolute minima for the ground-state JT APES. This derivation corroborates the intuition that the displacement $\mathbf{Q}_{N-1}$ provides maximal stabilization of the electronic ground-state, while simultaneously giving a very explicit illustration of the epikernel principle. Additional discussion of the latter is given in Sections III.C. and III.E.

\subsubsection{Topological equivalence between ground-state troughs and projective spaces}

In this section, we utilize the result obtained in the previous to demonstrate the topological equivalence (homeomorphism) between the space of minima of the ground-state APES of JT systems with continuous symmetries and the space of rays (lines) of real (for a spinless JT model) or quaternionic (in cases where spin-orbit coupling is strong) finite Hilbert spaces. The equivalence in the bosonic case has been previously pointed by Ceulemans \cite{ceulemans_a._structure_1987}. In any case, we provide a derivation for both bosonic and fermionic cases, as the latter result is a simple generalization of the former. Basically, we will show that the there exists a bijective relationship between the set of physically inequivalent electronic wave functions (Kramers doublets), i.e., unrelated by multiplication by a phase (Kramers pair rotation) in the spinless (fermionic) case, and the space of molecular geometries which are minima of the electronic JT APES. Topological equivalence follows from the continuity of the maps between the electronic states (doublets) and the minimal-energy molecular geometries.

To emphasize the remarkable physical content of the statement we are about to prove, we shall paraphrase the claim. It ascertains that an arbitrary electronic wave function (Kramers pair) is guaranteed to be the electronic (Kramers) ground state at a given molecular geometry in the minimum energy trough. Conversely, any molecular geometry in this trough is associated with a unique ground-state electronic wave function (Kramers), up to a phase (SU(2) rotation). Strikingly, even when all possible electronic wave functions are sampled along the trough, they correspond to the same electronic energy.

\emph{Bosonic troughs.} In accordance with the JT theorem, the adiabatic electronic ground-state of a molecular JT model is non-degenerate. As explained in the last sections, a continuous set $\mathcal{O}$ of global minima of the APES is guaranteed to exist in the presence of SO($N$) symmetry. In particular, if $\mathbf{Q}_0 \in \mathcal{O}$, then for any $\mathbf{Q}$ in $\mathcal{O}$, there exists $T(\mathbf{Q},\mathbf{Q}_0) \in \text{SO}(N)$, such that

\begin{equation} \mathbf{Q} = T(\mathbf{Q},\mathbf{Q}_0) \mathbf{Q}_0. \end{equation} 

In what follows, it will be important that at the space of ground-state minima $\mathcal{O}$, the electronic spectrum is assumed to be of the form explored in the previous section,
The most obvious implication of this spectrum is that the electronic Hamiltonian is invariant under unitary transformations of the electronic Hilbert space that act non-trivially only on the subspace spanned by the degenerate excited states. In particular, if $\ket{\psi_0(\mathbf{Q})}$ (with $Q \in \mathcal{O}$) denotes the non-degenerate electronic ground-state, and $U \ket{\psi_0(\mathbf{Q})} = \pm \ket{\psi_0(\mathbf{Q})}$, then $U \in \mathbb{Z}_2 \times O(N-1) \subset O(N)$, and $U H_{\text{JT}}(\mathbf{Q}) U^{-1} = H_{\text{JT}}(\mathbf{Q})$. The last equation implies that the rotation of the vibrational configuration space which corresponds to $U$ is the identity (see Sec. II and Eqs. \ref{hjts} and \ref{adjr}) .

We again follow the convention that the Hilbert space vectors corresponding to the adiabatic electronic ground-state at $\mathbf{Q} \neq 0$ are written as \textit{real} linear combination of electronic basis functions defined at the JT center, $\{\ket{\psi_1},\ket{\psi_2},...\ket{\psi_{N}}\}$. Thus, a normalized electronic ground-state wave function at $\mathbf{Q}_0 \in \mathcal{O}$ can be expressed as

\begin{align} \ket{\psi_0(\mathbf{Q}_0)} = & \sum_{i=1}^{N} c_{i0}(\mathbf{Q}_0) \ket{\psi_i}, c_{i0}(\mathbf{Q}_0) \in \mathbb{R}, \nonumber \\ & \sum_{i=1}^N c_{i0}^2(\mathbf{Q}_0) = 1.\end{align}

Thus, there exists a mapping of the ground-state trough to the $N-1$-dimensional sphere, $\mathbf{Q} \mapsto S^{N-1}$. However, the electronic ground-state wave function at $\mathbf{Q}_0 \in \mathcal{O}$ is only defined modulo a sign. In particular, if $\ket{\psi_0(\mathbf{Q}_0)}$ denotes a normalized eigenfunction of $H_{\text{JT}}(\mathbf{Q}_0)$ with lowest eigenvalue, then so is $-\ket{\psi_0(\mathbf{Q}_0)}$. Thus, the mapping $\mathcal{O} \rightarrow S^{N-1}$ is only well-defined locally (alternatively, it may be said to be \textit{double-valued}).  However, the electronic ground-state at $\mathbf{Q}_0$ can be defined unambiguously as a rank-one projection operator $\ket{\psi_0(\mathbf{Q}_0)}\bra{\psi_0(\mathbf{Q}_0)} \in \mathbb{R}P^{N-1}$. 
Hence, there exists a well-defined continuous function mapping the ground-state trough to the real projective space, $\text{$\Phi$:}~ \mathcal{O} \rightarrow \mathbb{R}P^{N-1}$, such that
 \begin{equation} \Phi(\mathbf{Q}_0) = \ket{\psi_0(\mathbf{Q}_0)}\bra{\psi_0(\mathbf{Q}_0)}. \end{equation}

Suppose $\mathbf{Q} = T^{-1}\mathbf{Q}_0, T \in \text{SO}(N)$, and let $U(T)$ denote a representative unitary transformation (e.g., let $U(T)$ be the unitary operator with $\text{det}[U(T)] = 1$) carrying the adiabatic ground-state at $\mathbf{Q}_0$ to that at $\mathbf{Q}\neq \mathbf{Q}_0$,
\begin{equation} U(T) \ket{\psi_0(\mathbf{Q}_0)} = \ket{\psi_0(T^{-1} \mathbf{Q}_0)}. \end{equation}
The map $\Phi$ satisfies the following equivariance condition
\begin{equation} \Phi(T^{-1} \mathbf{Q}_0) = U(T) \Phi(\mathbf{Q}_0) U^{-1}(T). \end{equation}
Thus, $\Phi$ is a continuous equivariant map of the electronic ground-state trough to $\mathbb{R}P^{N-1}$. $\Phi$ is also bijective. To show that it is injective, we assume $\Phi(\mathbf{Q}_0) = \Phi(\mathbf{Q}_1), \mathbf{Q}_0, \mathbf{Q}_1 \in \mathcal{O}$, and note that there exists $T \in \text{SO}(N)$, such that $\mathbf{Q}_1 = T^{-1} \mathbf{Q}_0$, since $\mathbf{Q}_0$ and $\mathbf{Q}_1$ belong to the ground-state trough. It follows that
\begin{equation} \Phi(\mathbf{Q}_0) = \Phi(T^{-1} \mathbf{Q}_0), \end{equation}
\begin{equation}  \ket{\psi_0(\mathbf{Q}_0)}\bra{\psi_0(\mathbf{Q}_0)} = U(T)\left(\ket{\psi_0(\mathbf{Q}_0)}\bra{\psi_0(\mathbf{Q}_0)}\right) U^{-1}(T). \end{equation}
The last equation is only satisfied if $U(T) \ket{\psi_0(\mathbf{Q}_0}) = \pm \ket{\psi_0(\mathbf{Q}_0)}$, so $U(T)$ acts non-trivially only the orthogonal subspace to the line spanned by $\ket{\psi_0(\mathbf{Q}_0)}$. This in turn implies $U(T) H_{\text{JT}}(\mathbf{Q}_0) U^{-1}(T) = H_{\text{JT}}(\mathbf{Q}_0)$, since we have assumed the excited-state spectrum to be degenerate. Hence, $T^{-1}\mathbf{Q}_0 = \mathbf{Q}_0$ and $\mathbf{Q}_1 = \mathbf{Q}_0$.  For the proof that $\Phi$ is surjective, let $p  \in \mathbb{R}P^{N-1}$. There exists a neighborhood of $p$,  $U_p \subset \mathbb{R}P^{N-1}$, on which a continuous section $\text{$\phi$:}~ U_p \rightarrow U_p \times S^{N-1}$ may be defined by
\begin{equation} \phi(p) = \sum_{i=1}^{N} c_{i}(p) \ket{\psi_i}. \end{equation}
Let $\mathbf{Q}_0$ be the configuration in the electronic ground-state trough for which $H_{\text{JT}}$ is diagonal [i.e., $\mathbf{Q}_0$ only has a non-vanishing component in the direction along which $H_{\text{JT}}(\mathbf{Q}_0)$ is diagonal (in the diabatic basis)] . There exists a lift of $\Phi(\mathbf{Q}_0)$ denoted by $\ket{\psi_0(\mathbf{Q}_0)}$, and an SO($N$) transformation $U(T_p)$, such that 

\begin{equation} \ket{p}_+ = U(T_p) \ket{\psi_0(\mathbf{Q}_0)} =  \ket{\psi_0(T_p^{-1}\mathbf{Q}_0)},\end{equation} 
where $T_p$ is the corresponding SO$(N)$ rotation of the vibrational configuration space. This follows from the continuous symmetry conditions, as is verified by

\begin{align} H_{\text{JT}}(T_p^{-1} \mathbf{Q}_0) \ket{p}_+ &=  H_{\text{JT}}(T_p^{-1} \mathbf{Q}_0) U(T_p) \ket{\psi_0(\mathbf{Q}_0)} \nonumber \\
& = \left[U(T_p)  H(\mathbf{Q}_0) U^{-1}(T_p) \right]U(T_p) \ket{\psi_0(\mathbf{Q}_0)} \nonumber \\ & = E_0(\mathbf{Q}_0) \ket{p}_+,
\end{align} 
where we used Eq. \ref{hjts}.
 Hence, $\Phi$ is shown to be a bijective continuous map between compact manifolds. Therefore, it is a homeomorphism \cite{lee2010introduction}.

\emph{Fermionic troughs.}
 In fermionic systems, Kramers degeneracy implies the JT distorted electronic ground-state  is a doublet $\ket{\Psi_0(\mathbf{Q}_0)} \equiv \{\ket{\psi_0(\mathbf{Q}_0)}, T \ket{\psi_0(\mathbf{Q}_0)}\}$ \cite{jahn_stability_1938}. Each normalized Kramers pair is only defined modulo multiplication by a unit quaternion $||q|| = 1$ (or equivalently, an SU(2) action) from the right (by convention). Given a definition of $N$ degenerate Kramers pairs $\ket{\Psi_i}$ at the JT center, the ground-state doublet at $\mathbf{Q} \neq 0$ may be written as
 
 \begin{equation} \ket{\Psi_0(\mathbf{Q})} = \sum_{m=1}^{N} \ket{\Psi_m}\cdot q_{0m}(\mathbf{Q}), q_{0m}(\mathbf{Q}) \in \mathbb{H}, \end{equation}
 where $q_{0m}(\mathbf{Q}) = \sum_{j=1}^{3} q_{0mj}(-i\sigma_j) + q_{0m4}\sigma_0$ is a quaternion (with imaginary units $-i\sigma_i$ and $2\times2$ identity matrix $\sigma_0$ representing the real generator) \cite{avron_chern_1989, levay_geometrical_1990}.
 If we require $\ket{\Psi_0(\mathbf{Q})}$ to be normalized, then
 \begin{equation} \braket{\Psi_0(\mathbf{Q})|\Psi_0(\mathbf{Q})} = \sum_m ||q_{0m}(\mathbf{Q})||^2 = 1, \end{equation}
where $||q||^2 = \text{det}(q^\dagger q)= q_0^2 + q_1^2 + q_2^2 + q_3^2$ is the quaternion norm. It follows that each normalized $\ket{\Psi_0(\mathbf{Q})}$ may also be viewed as a point on the sphere $S^{4N-1}$. 

Let $\mathbf{Q}_0$ belong to the ground-state trough $\mathcal{O}$. Corresponding to it there is an infinite number of equivalent ground-state Kramers pairs related by unit quaternion multiplication, $\{\ket{\Psi_0(\mathbf{Q}_0)}\cdot q\}$. The space of equivalence classes 

\begin{equation} \{ [\Psi] \sim [\Psi']~\text{if}~  \Psi' = \Psi\cdot q | \Psi, \Psi' \in \mathbb{H}^N, q \in \mathbb{H}^{*} \equiv \mathbb{H}-0\} \end{equation}
 of $\mathbb{H}^N$ defines the quaternionic projective space $\mathbb{H}P^{N-1} \cong \mathbb{H}^{N}/\mathbb{H}^*$ \cite{levay_geometrical_1990}. A non-degenerate ground-state (in the quaternionic sense) for all $\mathbf{Q} \in \mathcal{O}$, implies that there exists a well-defined map $\Phi$ between the adiabatic trough $\mathcal{O}$ and $\mathbb{H}P^{N-1}$. 
It may be explicitly given as
\begin{align} \text{$\Phi$:} ~& \mathcal{O} \rightarrow \mathbb{H}P^{N-1}, \nonumber\\ 
&\Phi(\mathbf{Q}_0) = \ket{\Psi_0(\mathbf{Q}_0)}\bra{\Psi_0(\mathbf{Q}_0)}.
\end{align}
This map is well-defined, since the projection operator is invariant under a redefinition of the basis, i.e., a gauge transformation, $\ket{\Psi} \mapsto \ket{\Psi}\cdot q$, with $q \in \mathbb{H}, ||q||^2 = 1$, as 
 $\ket{\Psi}\bra{\Psi} \rightarrow \left(\ket{\Psi}\cdot q\right)\left(q^{\dagger} \cdot \bra{\Psi}\right) = \ket{\Psi} ||q||^2 \bra{\Psi} = \ket{\Psi}\bra{\Psi}$. Moreover, continuous invariance under USp($2N)$ implies $\Phi$ satisfies an equivariance condition analogous to that discussed in subsection (a), 
\begin{equation} \Phi(T^{-1} \mathbf{Q}_0) = U(T) \Phi(\mathbf{Q}_0) U^{\dagger}(T), \end{equation}
except now $U(T)$ belongs to USp(2$N$). $\Phi$ can be shown to be bijective as in subsection (a). By compactness of $\mathcal{O}$ and $\mathbb{H}P^{N-1}, \Phi$ is a homeomorphism; this implies topological equivalence between the Kramers ground-state trough and $\mathbb{H}\text{P}^{N-1}$.

\subsubsection{Generic orbits}
The prior subsections provided a detailed description of the electronic ground-state equilibrium orbit of JT models carrying a maximal Lie group action. Here we will utilize the  continuous symmetry property of the studied models to obtain qualitative properties of their higher-energy orbits.

Let us consider first the spinless TRI case. The equivalence between rotations of the vibrational configuration space and special orthogonal transformations of the electronic basis implies the dimensionality of the space of JT distorted structures with a given adiabatic electronic spectrum is equal to or smaller than the dimensionality of SO$(N)$, $N = |\Gamma|$. It may be smaller because there may be matrices in SO$(N)$ which commute with $H_{\rm{JT}}(\mathbf{Q})$, in which case their action on the electronic Hamiltonian is equivalent to the identity matrix action on the vibrational configuration space, whence it follows that the nuclear configuration of the system remains invariant.

The subspace of $so(N)$ containing all matrices commuting with $H_{\rm{JT}}(\mathbf{Q})$ is called the \textit{centralizer} of $H_{\rm{JT}}(\mathbf{Q})$. We denote it by $\mathcal{N}(\mathbf{Q})$, 

\begin{equation} \mathcal{N}(\mathbf{Q}) = \left\{ M \in so(N) | [M,  H_{\rm{JT}}(\mathbf{Q})] = 0\right\}. \end{equation}
It can be checked that  $\mathcal{N}(\mathbf{Q})$ is a Lie subalgebra of $so(N)$ by applying the Jacobi identity,

\begin{align} [H_{\text{JT}},[N_1,N_2]] =  -[N_2,[H_{\text{JT}},N_1]] - [N_1,[N_2, H_{\text{JT}}]] = 0, \nonumber \\ ~\forall~ N_1, N_2 \in \mathcal{N}.\end{align}
Let $|\mathcal{N}(\mathbf{Q})|$ denote the dimensionality of $|\mathcal{N}(\mathbf{Q})|$. Then, the number of linearly independent JT distortions with the same adiabatic electronic spectrum $\{E(\mathbf{Q})\}$, which we denote by $|E(\mathbf{Q})|$, is given by 

\begin{equation} |E(\mathbf{Q})| = |so(N)| - |\mathcal{N}(\mathbf{Q})|. \label{dimreal} \end{equation}
Thus, the dimensionality of the space of molecular configurations with a given set of electronic energies depends on the number of degenerate eigenstates in the JT Hamiltonian when $\mathbf{Q} \neq 0$. This happens because a basis of $so(N)$ can always be constructed from antisymmetric matrices with only two non-vanishing entries, so that a diagonal matrix commutes with a basis vector of $so(N)$ iff the diagonal matrix elements which multiply the non-vanishing elements of the antisymmetric $so(N)$ matrices are equal. 

As a simple example, we now apply Eq. \ref{dimreal} to obtain the number of dimensions of the ground-state troughs of JT models with continuous symmetries. It is instructive to rederive this result, for the same method can be easily adjusted to infer te properties of higher-energy orbits. But, first, we need the dimensionality of the centralizer of $\mathbf{Q}$ at the trough, i.e., $|\mathcal{N}(\mathbf{Q})|$. It may be quickly obtained by making an assumption on the electronic spectrum at extremal regions of the JT APES, based on the epikernel principle \cite{ceulemans_symmetry_1984, ceulemans_a._structure_1987}. According to this, the stationary points of the ground-state APES are likely to be encountered in configuration space regions where the symmetry breaking of the molecular geometry is minimal. In the presence of the Lie group action discussed in this paper, this principle can be understood by considering the following. $\mathbf{Q}=0$ corresponds to a maximally symmetric molecular structure. All corresponding electronic eigenvalues are equal.  For any $\mathbf{Q} \neq 0$, the non-totally symmetric character of JT distortions implies the sum of the electronic eigenvalues $\lambda_i(\mathbf{Q})$ remains zero, i.e.,

\begin{equation} \sum_{i=1}^{N} \lambda_i(\mathbf{Q}) = 0. \end{equation}
A variety of possibilities exist for the splitting of the eigenvalues when $\mathbf{Q} \neq 0$, i.e., for the number of distinct degenerate multiplets which persist when the molecule undergoing JT effect is distorted along a given direction. We say that the molecular symmetry at $\mathbf{Q}=0$ is minimally broken at  $\mathbf{Q}_1$ if $\lambda_i(\mathbf{Q}) = x,~\forall~i ~\in \{1,2,...,N-1\}, x>0$ and $\lambda_N(\mathbf{Q}) = -(N-1)x$. This case is typical for the stationary points of JT systems with continuous symmetries \cite{obrien_dynamic_1969-2, pooler_jahn-teller_1977, ceulemans_so4_1989, chancey1997jahn}. It implies maximal stabilization of the adiabatic electronic ground-state, since all excited-states have their energy increased relative to the degenerate multiplet energy at $\mathbf{Q}=0$ (See Sec. III.B.1).

\begin{figure}
\includegraphics[width=0.45\textwidth]{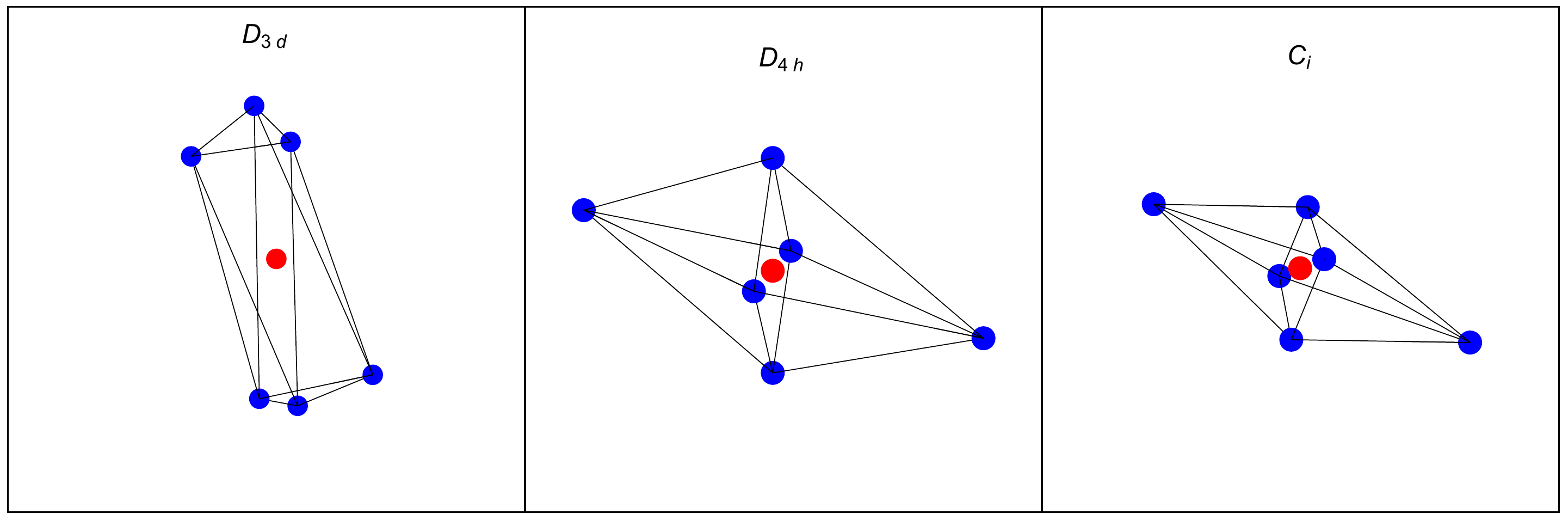}
\caption{$D_{3d}$, $D_{4h}$ epikernel and $C_i$ kernel structures for the $T_1 \otimes (e_g\oplus t_{2g})$ JT model. The first two have the same spectrum (in the presence of SO(3) symmetry) including a non-degenerate electronic ground-state and a doubly-degenerate excited state, while the latter is a lower symmetry structure with the same $r_1(\mathbf{Q})$ as the first two, but different $r_2(\mathbf{Q})$, reflecting the presence of three non-degenerate electronic states in its spectrum.}
\label{exepi}
\end{figure}

Based on the prior discussion, we now assume that for a minimal energy configuration $\mathbf{Q}_0$ of the JT system with continuous symmetry, only the lowest electronic energy eigenvalue is non-degenerate. The remaining $N-1$ are degenerate electronic excited states. Hence, a basis for $\mathcal{N}(\mathbf{Q}_0)$ consists of all matrices in $so(N)$ which mix the $N-1$ degenerate excited states at $\mathbf{Q}$ among themselves, but leave the ground-state invariant. This corresponds to the number of ways of arranging $(N-1)$ distinct indices into groups of two, i.e., $(N-1)(N-2)/2$. 
Hence, the dimensionality of the SO$(N)$-invariant adiabatic electronic ground-state trough is equal to

\begin{equation} |E(\mathbf{Q}_0)| = |\text{SO}(N)/\text{SO}(N-1)| = |S^{N-1}| = |\mathbb{R}P^{N-1}|,\end{equation}
\begin{equation} |E(\mathbf{Q}_0)| = \frac{N(N-1)}{2} -\frac{(N-1)(N-2)}{2} = N-1, \label{dimtrough} \end{equation}
which, of course, agrees with the arguments of the previous sections. 
The properties of higher-energy orbits can also be obtained with relative ease. The $T \otimes (e\oplus t_2)$ model with SO(3) invariance provides again a transparent example. Recall that this system has a two-dimensional orbit space, with each point labeled by $O(r_1,r_2)$ (Sec. IIIA), i.e., $O(r_1,r_2)$ is a submanifold of the $t_2 \oplus e$ vibrational configuration space, where the ground and excited-state JT APESs are flat. For a given $r_1 (\mathbf{Q}) = |\mathbf{Q}|^2  \neq 0$ the epikernel principle favors the minimal energy trough to be the 2D orbit $O(Q,1)$ by the following argument: the adiabatic electronic spectra for geometries in $O(Q,1)$ is given by

\begin{align} \text{spec}[H(O(Q,1))] = \{x,x,-2x\}, \nonumber \\ 
x=x(Q)>0, Q ~\in O(Q,1) \end{align}
while in $O(Q,r_2 \neq \pm 1)$  the JT Hamiltonian eigenvalues are 
\begin{align}&\text{spec}[H(O(Q,r_2))] = \{x',y',-x'-y'\},\nonumber \\ & x'=x'(Q,r_2)>0,~ x' > y'=y'(Q,r_2), r_2 \neq \pm 1. \end{align} 
Because $y'$ is intermediate between the lowest and highest-energy eigenvalues, its stabilization relative to the highest-energy eigenvalue of the spectrum at $O(Q,1)$ comes at the cost of an increased electronic ground-state energy (see Eq. 22). Hence, the ground-state trough is $O(Q_0,1)$, where $Q_0 = Q_0(\omega, F_v)$ minimizes the ground-state energy \cite{bersuker_jahn-teller_2006, obrien_dynamic_1969-2}. Section IIIA. showed $|O(Q_0,1)| = 2$ in accordance with application of Eq. \ref{dimtrough}. Note that because permutation of the eigenvalues can be effected by an SO(3) rotation, the assumptions made regarding the order of the eigenvalues do not imply loss of generality.
 
For any other model and type of degeneracy splitting of $H_{\text{JT}}(\mathbf{Q})$, we can apply the same methods and obtain the dimensionality and topology of the subspaces of the JT distorted structures with equal adiabatic electronic spectrum. For instance, if the electronic spectrum of $H_{\text{JT}}(\mathbf{Q})$ has a non-degenerate lowest-energy state and two degenerate doublets (such as it can happen in the icosahedral JT problem $H \otimes (g\oplus2h)$ \cite{ceulemans_a._and_fowler_r._jahnteller_1990}), the orbit is 8D according to

\begin{equation} |E(\mathbf{Q})| = \left|\frac{\text{SO}(5)}{\text{SO}(2)\times \text{SO}(2)}\right| = 8. \end{equation}

The above approach can be quickly adapted to models with strong spin-orbit coupling. The main difference is that the corresponding molecular Hamiltonians are globally invariant under the action of the unitary symplectic group USp($2N$) on spinorial and vibrational degrees of freedom. Further, independent USp(2)$\cong$SU(2) actions on each Kramers pair commute with the Hamiltonian (since they amount to a change of basis in a Kramers pair subspace). It follows that the number of dimensions of the ground-state trough of each spinorial JT model is given by:

\begin{equation} \begin{split} & |E(\mathbf{Q}_0)| = \left|\frac{\text{USp}(2N)}{\text{USp}(2N-2)\times \text{USp}(2)}\right| \\ & |E(\mathbf{Q}_0)| = N(2N+1)-(N-1)(2N-2) - 3 = 4(N-1), \label{krorbit}\end{split} \end{equation}
where we used that USp$(2N)$ is $N(2N+1)$-dimensional \cite{gilmore2012lie}.
As expected, this result agrees with the space of ground-state minima of the $\Gamma_8 \otimes (e\oplus t_2)$ ($N=2$) \cite{pooler_jahn-teller_1977} and $\Gamma_9 \otimes (g\oplus 2h)$ ($N=3$) \cite{ceulemans_jahn_2007} JT problems. The dimensionality of fermionic ground-state troughs bears a simple relationship to the spinless cases (cf. Eq. \ref{dimtrough}). It is a consequence of the quaternionic structure of the former. As seen in Sec. III.B.2, while the spinless trough is topologically equivalent to a real projective space \cite{ceulemans_a._structure_1987}, fermionic time-reversal symmetry implies topological equivalence of the Kramer's trough to the \textit{quaternionic projective space} $\mathbb{H}$P$^{N-1}$ \cite{levay_geometrical_1990}, the space of lines of $\mathbb{H}^{N}$. 

In the case of the icosahedral $\Gamma_9 \otimes (g\oplus 2h)$, another possibility for the JT splitting is given by choosing $\mathbf{Q}$ such that all three Kramers' pairs are non-degenerate. Then, the orbit space of the USp(6) action is 12-dimensional since it can be parametrized by $\text{USp(6)}/\text{USp(2)}\times \text{USp(2)} \times \text{USp}(2)$. Here we find an interesting parallel between $\Gamma_9 \otimes (g\oplus 2h)$ and $T \otimes (e\oplus t_2)$: in both cases the orbit space is two-dimensional. The ground-state trough belongs to a 2D orbit in $T \otimes (e\oplus t_2)$, while it lives on an 8D orbit in the case of $\Gamma_9 \otimes (g\oplus2h)$ (Sec. IIIB). Except for a set of measure zero, higher-energy orbits of $T \otimes (e\oplus t_2)$ and $\Gamma_9 \otimes (g\oplus2h)$ are 3D and 12D, respectively. Thus, the number of dimensions of the orbits of $\Gamma_9 \otimes (g\oplus2h)$ is always a factor of 4 larger than those of $T \otimes (e\oplus t_2)$, which generalizes Eq. \ref{krorbit}.

\subsection{Geometric phase and ground-state degeneracy}

The occurrence of the geometric phase \cite{berry_quantal_1984} in JT and related models has a long history \cite{longuet-higgins_studies_1958, obrien_dynamic_1969-2, zwanziger_topological_1987, ham_berrys_1987, romerorochin_time_1989, ceulemans_berry_1991, auerbach_vibrations_1994, cullerne_jahn-teller_1994,  de_los_rios_dynamical_1996, bersuker_jahn-teller_2006, requist_molecular_2016-1, ryabinkin_geometric_2013}. It is highly relevant in the strong vibronic coupling limit, where $F_\Lambda^2/(2 k_\Lambda) >> 1$. In particular, it provides an unambiguous explanation for the non-generic feature (in finite physical systems) of vibronic ground-state degeneracy in some JT models \cite{bersuker_jahn-teller_2006,chancey1997jahn}. In this section, we discuss the Berry phases of JT models carrying a Lie group action. 

The abelian Berry phase \cite{berry_quantal_1984} characterizes the twisting of a family of 1D complex vector spaces parametrized by a configuration space $C$ (a line bundle). It can be computed in a variety of ways \cite{berry_quantal_1984, ceulemans_berry_1991, mead_geometric_1992}. The most significant aspect to our discussion is that for time-reversal invariant bosonic systems, the adiabatic geometric phase belongs to $O(1) = \mathbb{Z}/2\mathbb{Z} = \{+1,-1\}$. This is a consequence of the existence of a real structure in the Hilbert line bundle (family of 1D Hilbert spaces) when $T^2=1$. In other words, a real locally continuous choice of eigenstates of a parameter-dependent Hamiltonian may be chosen for each point of the parameter space, so only $\pm 1$ Berry phases are allowed.

In the case of fermionic systems Kramers' degeneracy implies the existence of pairs of degenerate states. A normalized pair is only defined modulo an SU(2) transformation, or equivalently, quaternion multiplication \cite{avron_chern_1989,levay_geometrical_1990}. As a result, the geometric phase of the ground-state of a gapped fermionic system is a (non-abelian) unit quaternion (or SU(2) matrix in the complex representation)\cite{avron_chern_1989,levay_geometrical_1990,mead_molecular_1987}.

JT models carrying a Lie group action are particularly susceptible to having relevant Berry phase effects. This happens because, as described in the previous subsections, the vibrational configuration space can be decomposed into topologically non-trivial orbits $O_G$ of a Lie group $G$, which may admit twisted families of electronic Hilbert spaces, due to the existence of level crossing at the JT center.

\subsubsection{Bosonic models}

For spinless systems, the electronic ground-state of $\hat{H}_{\text{JT}}(\mathbf{Q}), ~\text{with}~ \mathbf{Q} \neq 0$, is non-degenerate in the minimal-energy JT trough. It is homeomorphic to the space of lines in the $N$-dimensional real vector space, the real projective space $\mathbb{R}P^{N-1}$ (see Sec. III.B.2).  The sphere $S^{N-1}$ is a double cover of $\mathbb{R}P^{N-1}$, i.e., $\mathbb{R}P^{N-1} \cong S^{N-1}/\mathbb{Z}_2$. Therefore, if $N > 2$, it has two equivalence classes of loops, as $\pi_1(\mathbb{R}P^{N-1}) = \mathbb{Z}_2$ ($\mathbb{R}P^1$ is isomorphic to $S^1$, which has its loop classes distinguished by an integer, the winding number, so $\pi_1(S^1) = \mathbb{Z}$)\cite{nakahara2003geometry}. The identity is given by the class of closed curves on $S^{N-1}$ (as it is simply-connected when $N > 2$). The other element of $\pi_1(\mathbb{R}P^{N-1})$ can be represented by an open continuous curve on $S^{N-1}$ connecting antipodal points of the latter, which are the same on $\mathbb{R}P^{N-1}$. The product of two non-trivial (odd) loops gives a trivial one, but a trivial (even) path followed by a non-trivial is equivalent to the latter. Loops in the even class do not give a Berry phase for any quantum state since they can be continuously deformed into the constant path, which obviously, has no Berry phase. Conversely, the odd loops of $\mathbb{R}P^{N-1}$ \textit{may} give rise to a nontrivial geometric phase for a non-degenerate adiabatic electronic state. In particular, this will happen if the line bundle associated to the non-degenerate electronic state is twisted, i.e., if a continuous global family of \textit{real} adiabatic electronic wave functions $\mathbf{Q} \mapsto \psi_0(\mathbf{Q})$ does not exist. In the case of $\mathbb{R}P^1$, we may group all loops with even winding number in the trivial class, since they give rise to no geometric phase. Odd loops correspond to those with odd winding number. They give rise to a Berry phase in the case of the $E \otimes e$ model, as illustrated by Fig. \ref{exepseudo}.

If a Berry phase exists for a given \textit{real} adiabatic electronic state, the vibrational wave functions are required to satisfy antiperiodic boundary conditions for the molecular wave function to be single-valued. These cannot be satisfied by vibrational wave functions transforming in the totally symmetric irrep of SO$(N)$, for they are even under inversion \cite{barut1986theory}.  All other irreps of SO$(N)$ are degenerate.  Therefore, the vibronic ground-state will be degenerate whenever the adiabatic electronic ground-state admits a Berry phase.
In particular, it will belong to the vector irrep of SO$(N)$ when $N > 2$ \cite{obrien_dynamic_1969-2, auerbach_vibrations_1994, cullerne_jahn-teller_1994, chancey1997jahn, de_los_rios_dynamical_1996}. This is the case, for the Schrodinger equation describing the pseudorotational motion in the adiabatic electronic ground-state trough has the same form as that for particle motion on an $N-1$-dimensional sphere (with the equivalence of antipodal points under the mapping to the real projective space enforced by a boundary condition), for which the set of lowest-energy eigenstates that are odd under inversion about $\mathbf{Q}=0$ belongs to the vector irrep of SO($N$)\cite{obrien_dynamic_1969-2, cullerne_jahn-teller_1994, de_los_rios_dynamical_1996, chancey1997jahn}.

The above picture has been shown to break down in the presence of warping of the ground-state JT APES. In a variety of examples \cite{zwanziger_topological_1987, koizumi_multiple_2000,  chancey1997jahn, lijnen_berry_2005}, it was found that the inclusion of sufficiently strong anisotropy in the ground-state APES resulted in the formation of a set of esis which are circulated by low-energy vibrational tunneling paths on the electronic ground-state APES. This, in turn, changes the Berry phase of the adiabatic electronic states, and modifies the symmetry of the vibronic ground-state. Thus, a rearrangement of vibronic energy levels (ground-state crossover) is seen as a function of quadratic vibronic coupling, or of difference in JT stabilization energies for vibrational modes in different irreps of the molecular point group at $\mathbf{Q} =0$. 

For JT systems with maximal continuous invariance under the action of a Lie group, no such ground-state crossover can happen. This has been verified for each individual case with $T^2=1$ of Table 1 \cite{longuet-higgins_studies_1958, obrien_dynamic_1969-2, cullerne_jahn-teller_1994, de_los_rios_dynamical_1996, chancey1997jahn}, and it may be understood in the following way: for any $\mathbf{Q}$ in the adiabatic electronic ground-state trough there exists a basis for which the $N$-dimensional electronic JT Hamiltonian is diagonal with eigenvalues $\{x,x,...,x,-(N-1)x\} $, where $x > 0$. A closed path on the ground-state trough can be parametrized by SO($N$) rotations of the vibrational configuration space, which leave the adiabatic electronic spectrum invariant. By checking the phase acquired by the adiabatic ground-state upon a non-trivial loop of $\mathbb{R}P^{N-1}$, it has been shown that it \textit{necessarily} changes sign\cite{longuet-higgins_studies_1958, obrien_dynamic_1969-2, cullerne_jahn-teller_1994, de_los_rios_dynamical_1996, chancey1997jahn}. Because the existence and properties of the trough are independent of the reduced vibronic coupling constant and vibrational frequency, the Berry phase in the spinless models of Table 1 is robust with respect to both changes in the fundamental parameters of these models which preserve their fundamental symmetry, and perturbations that break the symmetry, but do not induce ground-state esis in low-energy regions of the JT APES, besides that at the JT center.

\subsubsection{Fermionic models}

In models with strong spin-orbit coupling, $H_{\text{JT}}(\mathbf{Q})$ remains degenerate even when $\mathbf{Q} \neq 0$, but this ceases to be true in the quaternionic representation of the corresponding complex Hilbert space (where a quantum state and its time-reversal partner define a single vector) \cite{jahn_stability_1938, avron_chern_1989}. As derived in Sec. III.B.3, in spinorial models with continuous symmetries, the ground-state trough is homeomorphic to the quaternionic projective space $\mathbb{H}P^{N-1} \equiv \mathbb{H}^{N}/\mathbb{H}-\{0\}$. The normalized ground-state Kramers' doublet is only defined modulo a local $su(2)\cong usp(2) \cong \mathbb{H}$ transformation. The impossibility of assigning a global continuous parametrization of the ground-state Kramers' doublet for all $\mathbf{Q} \in \mathbb{H}P^{N-1}$ implies the existence of a non-abelian (or, in this case, a quaternionic) Berry phase \cite{mead_molecular_1987, avron_topological_1988}, as it is transported according to the adiabatic theorem. 

To each ground-state Kramers pair, there corresponds a quaternionic line bundle, which completely characterizes the corresponding allowed nonabelian Berry phases. While the classification of quaternionic line bundles over $\mathbb{H}P^N$ for general $N$ is unknown, solutions exist for the JT models of interest to our study, (a) $\Gamma_8 \otimes (e\oplus t_2)$ ($N=1$) \cite{pooler_jahn-teller_1977} and (b) $\Gamma_9 \otimes (g\oplus 2h)$ ($N=2$) \cite{ceulemans_jahn_2007}. In case (a) the spinorial irreps can be embedded into the $J=3/2$ irrep of SU(2), and the active JT modes are quadrupolar (they form a basis for the $J=2$ irrep of SU(2)) \cite{ceulemans2013group}. The corresponding spinor JT Hamiltonian is equivalent to that thoroughly investigated by Avron et al. \cite{avron_chern_1989}. They showed that only two topologically distinct quaternionic line bundles exist for this model, with second Chern number \cite{nakahara2003geometry} equal to $\pm 1$ \cite{avron_topological_1988}. The non-abelian Berry phase corresponds to an SU(2) transformation, as expected. This implies a much richer set of behaviors for Kramers' partners undergoing cyclic adiabatic evolution as the nuclear geometry pseudorotates, including e.g., the possibility of population inversions \cite{mead_molecular_1987, avron_chern_1989, apsel_berry_1992}.

The spinorial Hamiltonian of $\Gamma_9 \otimes (g\oplus 2h)$ has only recently been constructed \cite{ceulemans_9_2007}. The electronic states form a basis for the $J=5/2$ irrep of SU(2), while the JT modes $g\oplus h$ can be embedded into the hexadecapolar $(J=4)$ irrep of SU(2), and the remaining $h$ vibrations are quadrupolar ($J=2$). The extrema of the ground-state APES generate an 8D trough isomorphic to the quaternionic projective plane $\mathbb{H}P^2$. Unfortunately, while the isomorphism $\mathbb{H}P^1 \cong S^4$  allows for a simplified analytical treatment of $\Gamma_8 \otimes (e\oplus t_2)$, the properties of $\mathbb{H}P^2$ are much more complex (though the classification of its quaternionic line bundles has also been accomplished \cite{gonccalves2006quaternionic}). For instance, the second Chern class \cite{avron_topological_1988} may be utilized to classify its 4D submanifolds, but it is not sufficient to uniquely identify the quaternionic line bundles over $\mathbb{H}P^2$ (in contrast with the $\mathbb{H}P^1$ case, for which all quaternionic line bundles are completely characterized by the second Chern class)\cite{avron_topological_1988}. A simpler situation arises if the hexadecapolar distortions are disallowed. Then, the spinor JT Hamiltonian becomes isomorphic to that of a $J=5/2$ spin coupled to an external quadrupole field. The topological classification of the quaternionic line bundles over the space of Hamiltonians of this class was given in Ref. \cite{avron_chern_1989}.

\subsection{Symmetry breaking}

In this subsection we provide brief comments on the significance of the properties of JT models carrying a Lie group action when their continuous symmetry is broken.

We have already noted that the APESs become warped in the presence of higher-order vibronic couplings and/or anisotropic JT stabilization energies/vibrational frequencies in the multimode problem. By the epikernel principle\cite{ceulemans_a._structure_1987,ceulemans_epikernel_1989}, while a non-degenerate ground-state results from the static JT symmetry breaking, a subset of the remaining APESs are likely to be degenerate at equilibrium positions of the distorted system. The possible epikernels represent different ways to break the adiabatic electronic state degeneracy at the JT center, while preserving some non-trivial subgroup of the point group defining the JT model. This was explored in previous subsections. We also showed that radial and angular coordinates can be employed to characterize the splitting of the Lie group-invariant JT APES. In particular, each set of radial displacements specifies an orbit, the internal space of which includes molecular structures with the same electronic spectrum, that can interconvert via pseudorotation. There exists as many distinct radial coordinates as linearly independent ways to split the degenerate multiplet. For example, in $T \otimes (e\oplus t_2)$, the radial coordinates specify whether a given nuclear geometry has a degenerate electronic subspace.  Higher-rank epikernels correspond to molecular deformations along coordinates which remove the degeneracy at the JT center minimally (e.g., by letting $N-1$ of the branches of the JT PES to remain degenerate). On the other hand, pseudorotational motion provides distortions which continuously map a given epikernel distortion into another that is equivalent. However, in a warped APES, pseudorotation is hindered, due to the formation of energetic barriers between the discrete set of minima. Yet, as shown by Ceulemans \cite{ceulemans_a._structure_1987}, coordinates adapted to a continuous group action provide a simple description of the set of extrema of the warped APES.

\section{JT models with continuous symmetries and generic conical intersections}

In this section we demonstrate a common feature of molecular accidental cis \cite{herzberg_intersection_1963, stone_spin-orbit_1976, worth_beyond_2004} and JT systems with maximal continuous symmetries. The latter satisfy the following condition: the number of JT active displacements is \textit{equal} to the minimal required for the intersection of a given number of APESs to happen at isolated points of the configuration space (without any assumed symmetry). We show below there exists a correspondence between accidental cis and the aforementioned JT systems with electronic multiplets consisting of $2-5$ intersecting states (the SO(2), SO(3), SO(4), and SO(5) models) when spin-orbit coupling is irrelevant, and 4 and 6 states (the USp(4) and USp(6) models) when spin-orbit coupling is strong. While in this section we rederive a well-known result on the co-dimension of parameter spaces admitting accidental cis \cite{von_neumann__1929}, this is done in a way that emphasizes the similarity between JT models with Lie group invariance and the behavior of a molecular system in the neighborhood of accidental cis. The significance of this result is outlined below.

Let us start with the case where $T^2=1$: suppose $N$ states are degenerate at a point $\mathbf{Q} = 0 $, $\mathbf{Q} \in \mathbb{R}^{P_N}$, where $P_N$ is equal to the minimum number of parameters which need to be tuned to generate an $N$-state conical intersection. Let  $E_1(0) = E_2(0) = ... = E_N(0) = 0$. Then, in a small neighborhood around the conical intersection,

\begin{equation} H(\mathbf{Q}) = \sum_{i=1}^{\rm P_N} Q_i M_i, \end{equation}
where we assume the $M_i$ are traceless symmetric matrices. This implies $iM_i$ belongs to the $su(N)$ Lie algebra, which has dimensionality $|su(N)| = N^2-1$ \cite{pfeifer2003lie}. 

We can always choose a basis for $su(N)$  where $N(N+1)/2 -1$ basis vectors are real-symmetric matrices while the remaining $N(N-1)/2$ are real-antisymmetric \cite{pfeifer2003lie}. The latter provide an embedding of $so(N)$ into $su(N)$. The $M_i$ do not belong to $so(N)$. Therefore, the number of independent matrices which can be employed in the linearization of $H(\mathbf{Q})$ around the conical intersection point is equal to $N(N+1)/2-1$, i.e.,

\begin{equation} P_N = \frac{N^2+N-2}{2}. \end{equation}
Note $P_1 = 0, P_2 = 2, P_3 = 5, P_4 = 9, P_5 = 14$. These are the number of free parameters which are required to vanish in order to have isolated degeneracies of $1,2,3,4$ and $5$ states in asymmetric systems. They agree with the number of JT active vibrational modes in each of the models of Table 1.

The case where $T^2=-1$ requires the electronic Hilbert space to be even-dimensional. Thus, the possible $iH(\mathbf{Q})$ form a subspace of the $su(2N)$ Lie algebra. The latter has $4N^2-1$ generators. However, the matrices in the subspace $usp(2N) \subset su(2N)$ may not be used as building blocks of $iH(\mathbf{Q})$, since $H(\mathbf{Q})$ is required to be hermitian. Hence, the dimensionality of the space of fermionic TRI Hamiltonians is given by

\begin{equation} P_N = |su(2N)|-|usp(2N)| = N(2N -1)-1.\end{equation} 

For $N=2$ and $3$ we obtain the number of coordinates involved in the JT distortions of $\Gamma_8 \otimes (e\oplus t_2)$ and $\Gamma_9 \otimes (g\oplus2h)$, respectively.

Just like $E \otimes e$ is a paradigmatic model for cis of two states in the presence of time-reversal symmetry and no spin-orbit coupling \cite{longuet-higgins_studies_1958, herzberg_intersection_1963, longuet-higgins_intersection_1975}, we have demonstrated that the remaining JT models with continuous symmetries could also be deemed as canonical models of generic behavior near cis in cases where more than two branches of a PES intersect. They are indeed the simplest models where these appear.

The perspectives introduced by this view have at least two reasons for being relevant to the treatment of dynamics near generic cis: (i) from a practical perspective the APES near a level crossing shows the same simplifying features as the JT models carrying a Lie group action, e.g., pseudorotational directions along which the electronic spectrum is invariant; (ii) from a conceptual point of view, molecular distortions which lift the degeneracy at cis can be assigned a qualitative meaning based on the correspondence with a JT model, e.g., motions near a triplet conical intersection can be classified in terms of quadrupoles, according to their effects on the electronic spectrum and the analogy with the SO(3)-invariant $T \otimes (e\oplus t_2)$ model. 

\section{Epilogue}

We have (i) shown that the space of electronic Hamiltonians of certain JT models invariant under a Lie group action on electronic and vibrational degrees of freedom can be identified with symmetric spaces in the orthogonal SU($N$)/SO($N$) and symplectic classes SU($2N$)/USp($2N$) in the spinless and fermionic cases, respectively; (ii) employed the invariance properties of the symmetric spaces to substantially reduce the complexity of the adiabatic electronic spectra of these high-dimensional models, by showing they can be decomposed into orbit spaces of SO$(N)$ or USp$(2N)$ (in the spinless and fermionic models, respectively), and identified motion within each orbit with molecular pseudorotation; (iii) explained the relationship between these constructions and the abelian and non-abelian Berry phases of these models; (iv) discussed the qualitative significance of the studied models when the continuous symmetry is broken, i.e., for more realistic systems, and (v) demonstrated that a common characteristic of the investigated systems is that they include the minimal number of parameters required to induce accidental cis involving up to 5 states of spinless models, and 6 states when spin-orbit coupling is strong. Thus, some of their generic features are expected to be present in systems exhibiting cis involving many states, but no molecular symmetry. We have also provided novel derivations for the topological equivalence between the ground-state trough of JT models and the real and quaternionic projective spaces in the spinless and fermionic models, respectively, and proved that the electronic adiabatic spectrum of the continuous set of minima of all JT models with maximal continuous symmetries destabilizes all states, except for the lowest. Several questions remain to be explored, e.g., we have avoided any discussion related to non-adiabatic dynamics \cite{yonehara_fundamental_2012, matsika_nonadiabatic_2011}, even though the corresponding couplings can be determined straightforwardly for the systems studied with the employed methodology. The physical implications of the topological classification of fermionic systems with strong spin-orbit coupling and more than two intersecting Kramers pairs also remain to be understood. Models with continuous symmetries such as those discussed here undoubtedly provide the simplest platform for any study of these and related matters

\section{Acknowledgments} We acknowledge NSF CAREER award CHE: 1654732 and generous UCSD startup funds.

\bibliography{lib}
\end{document}